\documentclass{llncs} 
\pdfoutput=1

\def\R{{\mathbb R}}
\newcommand{\scsc}{\textsf{SC}$^2$ }
\usepackage[figure,table]{totalcount}

\usepackage{xspace}
\usepackage{booktabs}
\usepackage{url}
\usepackage[show]{ed}
\usepackage{amsmath}
\usepackage{parskip}
\usepackage{hyperref}
\usepackage{commath}
\usepackage{amsfonts}
\usepackage{amssymb}
\usepackage{graphicx}
\usepackage{wrapfig}
\usepackage{xcolor}
\usepackage[utf8]{inputenc}
\usepackage{listings}
\usepackage{color}

\usepackage[ruled]{algorithm}
\usepackage{algorithmic}[1]

\newcommand{\head}[1]{\textnormal{\textbf{#1}}}
\usepackage[english]{babel}

\begin{document}

\pagenumbering{roman}
\setcounter{page}{1}
\mainmatter

\title{Towards Incremental \\Cylindrical Algebraic Decomposition}
\subtitle{Report on Summer Project for the \scsc Project}
\titlerunning{Towards Incremental Lazard Cylindrical Algebraic Decomposition}  
%
\author{Alexander I. Cowen-Rivers\inst{1} \\ Supervisor: Matthew England\inst{2}}
\authorrunning{A.I. Cowen-Rivers and M. England} 
%
%
\institute{
Department of Computer Science, \\
University College London, London, UK \\
\email{a.cowen-rivers.17@ucl.ac.uk}
\and
Faculty of Engineering, Environment and Computing, \\
Coventry University, Coventry, UK \\
\email{Matthew.England@coventry.ac.uk}
}

\maketitle              

\begin{abstract}

Cylindrical Algebraic Decomposition (CAD) is an important tool within computational real algebraic geometry, capable of solving many problems to do with polynomial systems over the reals, but known to have worst-case computational complexity doubly exponential in the number of variables.  
It has long been studied by the Symbolic Computation community and is implemented in a variety of computer algebra systems, however, it has also found recent interest in the Satisfiability Checking community for use with SMT-solvers.  
The \scsc Project seeks to build bridges between these communities.  

The present report describes progress made during a Research Internship in Summer 2017 funded by the EU H2020 \scsc CSA.  We describe a proof of concept implementation of an Incremental CAD algorithm in Maple, where CADs are built and refined incrementally by polynomial constraints, in contrast to the usual approach of a single computation from a single input.  This advance would make CAD of use to SMT-solvers who search for solutions by constantly reformulating logical formula and querying solvers like CAD for whether a logical solution is admissible.
We describe experiments for the proof of concept, which clearly display the computational advantages when compared to iterated re-computation.   

In addition, the project implemented this work under the recently verified Lazard projection scheme (with corresponding Lazard evaluation).  That is the minimal complete CAD method in theory, and this is the first documented implementation.
\end{abstract}

\newcounter{Count}
\newcounter{Section}

\newpage

\section{Introduction}


The aim of this project is to adapt the main Cylindrical Algebraic Decomposition (CAD) algorithm for use with SMT-solvers, as part of the \scsc Project which seeks to build collaborations between researchers in Symbolic Computation and Satisfiability Solving \cite{AAB+16a}.  The work is influenced by the success of SAT/SMT solvers while conducting Conflict-Driven Clause Learning (CDCL) search based algorithms (see for example \cite{BHvMW09}), and their recent application to the domain of non-linear real arithmetic where CAD provides the only complete theory.

The main contributions of this report are the description of algorithms for adapting a CAD to changes in the input it is built with respect to: either incrementing or reducing the polynomial system by one constraint.  We also describe results for a proof of concept implementation in the computer algebra system Maple.  This shows promising experiment results with savings of up to half the calculation time, compared to a full CAD recalculation. 


Other contributions are on the application side of Symbolic Computation including: an implementation of the Lazard projection and evaluation.  The Lazard projection operator was proposed in 1994 \cite{lazard_1994}, but shortly after a flaw was found in its proof of correctness (see \cite{MH16} for details).  However, recent work \cite{MPP17} has given an alternative proof (which necessitates changes to the lifting stage) renders this the most efficient complete CAD projection operator.  

\par 

\subsection{Terminology}

First, we will introduce some of the basic definitions needed to follow this article.  We follow the presentation in the lecture notes by Jirstrand, \cite{jirstrand_1995}.  We work over $n$-dimension real space $\mathbb{R}^n$ in which there is a variable ordering $x_1 \prec x_2 \prec \dots \prec x_n$.

\textbf{Definition \stepcounter{Count}\arabic{Count}} A \textit{decomposition} of the space \(\mathcal{X} \subset \mathbb{R}^n\) is a finite collection of disjoint regions (components) whose union is \(\mathcal{X}\).
\par 

\textbf{Definition \stepcounter{Count}\arabic{Count}} A set is \textit{semi-algebraic} if it can be constructed by finitely many applications of $union$, $intersection$ and $complementation$ operations on sets of the form \(\{ x \in \mathbb{R}^{n}\ |\ \mathbf{f}(x) \geq 0 \}\) where \(\mathbf{f} \in  \mathbb{R}[x_{1},\cdots,x_{n}]\).

\par 

\textbf{Definition \stepcounter{Count}\arabic{Count}} A decomposition \(\mathcal{D}\) is \textit{algebraic} if each of its components \( x \in \mathcal{D}\) is a semi-algebraic set.
\par 

\textbf{Definition \stepcounter{Count}\arabic{Count}} A finite partition of \(\mathcal{D}\) of \(\mathbb{R}^n\) is called a \textit{cylindrical decomposition} of \(\mathbb{R}^n\) if the projections of any two cells onto any lower dimensional coordinate space with respect to the variable ordering are either equal or disjoint.

\par 
Note: any cylindrical decomposition of \(\mathbb{R}^{n}\) implies additional cylindrical decompositions of \(\mathbb{R}^{n-1}, \ldots, \mathbb{R}^{1}\).   
\par 

\textbf{Definition \stepcounter{Count}\arabic{Count}} If a cylindrical decomposition is also an algebraic decomposition, then it is a \textit{cylindrical algebraic decomposition} (CAD). 
\par 

\textbf{Definition \stepcounter{Count}\arabic{Count}} A CAD is traditionally produced \textit{sign-invariant} with respect to a set of input polynomials, which means that each polynomial has constant sign (positive, negative or zero) on each cell.

\textbf{Definition \stepcounter{Count}\arabic{Count}} The elements of a CAD are refereed to as \textit{cells}. Traditionally, each cell is equipped with: a \textit{cell index} which is a list of integers which defines the position of a cell in the decomposition (first index refers to $x_1$ etc.); and a \textit{sample point} of the cell.  The cells we produce also come with a \textit{cell description}: a cylindrical formula.   

For example, in a CAD \(\mathcal{C} \in \mathbb{R}^{2}[x,y]\), a cell could be defined by the triple 
\(
[[1,1],[x<1,y<1],[0,0]]
\)
where: the first element \(([1,1])\) specifies the index of the cell (identifying it as the least with respect to both dimensions); the second element \(([x<1,y<1])\) gives the exact description of the cell; and finally the third element \(([0,0])\) specifies a sample point within the cell. 
\par 

\subsection{Example}

\begin{figure}[H]
  \centering
  \begin{minipage}[b]{0.48\textwidth}
      
      \includegraphics[width=.95\linewidth]{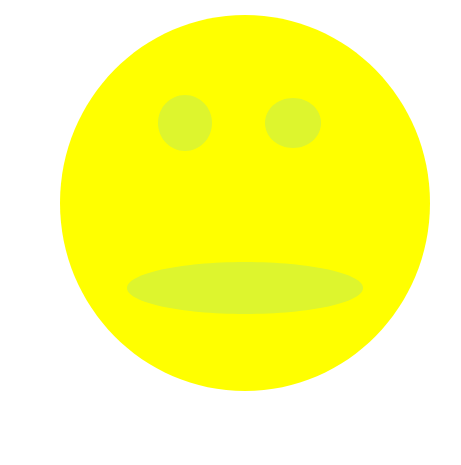}
    \caption{Closed curve reduction}
    \label{fig:ging}
  \end{minipage}
  \hfill
  \begin{minipage}[b]{0.48\textwidth}
      \includegraphics[width=.95\linewidth]{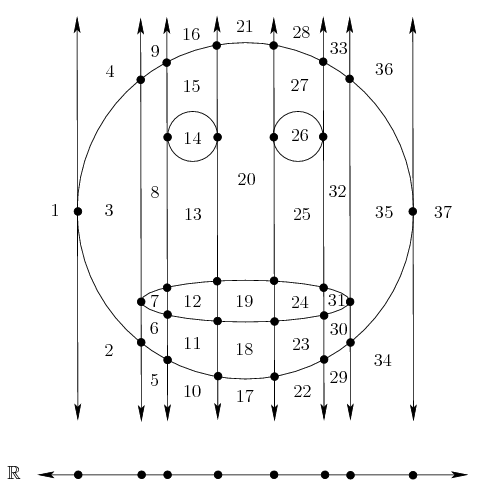}
    \caption{A CAD of Fig \ref{fig:ging}. }
        \label{fig:ging2}

  \end{minipage}
\end{figure}

We give a visual example inspired by \href{url}{http://planning.cs.uiuc.edu/node296.html}.  We first take a gingerbread face and remove the detail to leave an image resembling Figure \ref{fig:ging}.  This is formed by four closed curves, each of which will be defined by a bi-variate polynomial equation.

A CAD on this system of polynomials is visualised in Figure \ref{fig:ging2}.
Notice there are $37$ open cells (those of two dimensions).  There are a further $28$ partially open (1-dimensional line segments) and $28$ closed cells (isolated points) giving $93$ CAD cells in total.

It is worth noting that in many industrial applications (especially those from SMT applications), the polynomial systems will not form aesthetically pleasing geometric shapes, and will often better resemble randomly generated systems.

\subsection{New objects of study}

We now formally define the two main objects of study in this report.
 
\textbf{Definition \stepcounter{Count}\arabic{Count}} An \textit{InCAD} of the space 
$\mathbb{R}^n$ is an incremental cylindrical algebraic decomposition, caused by adding a polynomial $f_{new}$ into a polynomial system $\{f_1, \dots, f_m\} \in \mathbb{R}[x_1, \dots ,x_2]$, for which there was already has a CAD computed, producing a new CAD of $\{f_1, \dots, f_m, f_{new}\} \in \mathbb{R}[x_1,\dots,x_n]$.

\textbf{Definition \stepcounter{Count}\arabic{Count}} A \textit{ReCAD} of the space 
$\mathbb{R}^n$ is a reduced cylindrical algebraic decomposition, caused by removing a polynomial $f_{m}$ from $\{f_1, \dots, f_m\} \in \mathbb{R}[x_1\, \dots,x_n]$, which already has the  CAD computed, producing a new CAD of $\{f_1,..f_{m-1}\} \in \mathbb{R}[x_1,\dots,x_n]$.

\subsection{Report plan}

The report proceeds as follows.  In Section 2 we describe Lazard's Projection scheme and in Section 3 the development and results of an incremental projection algorithm built upon this.  Then in Section 4, we describe the lifting stage necessitated by Lazard projection and in Section 5 the development and results of an incremental lifting algorithm via the Lazard evaluation.  We finish in Section 6 with a summary of possibilities for future work.

\newpage

\section{Projection}

\subsection{Lazard projection}

The present project built upon code from the open source \textsc{ProjectionCAD} package \cite{EWBD14} for \textsc{Maple}.  This implemented the McCallum family of projection operators following \cite{McCallum1998} and needed to be adapted for the Lazard projection scheme.

First, we will introduce some notation. When we speak of projection, we usually refer to the maximal chain of projections. I.e. if a single projection eliminates one variable
\[
\pi_{n-1} : \mathbb{R}^n \rightarrow \mathbb{R}^{n-1} : (x_n,\ldots,x_1) \mapsto (x_{n-1},\ldots,x_1)
\]
then we may consider chaining these with the maximal chain one less than the total number of variables 
\[
\pi^{*} : \mathbb{R}^n \rightarrow \mathbb{R}^{1} : (x_n,\ldots,x_1) \mapsto (x_1).
\]
\par 

In this article we will refer to the Lazard projection operator \cite{lazard_1994}, \cite{MPP17} as \textit{ProjL} and the McCallum projection operator \textit{ProjM} \cite{McCallum1998}. 

Algorithm \ref{alg:1} shown later performs the Lazard projection.  It is almost identical to the counterpart McCallum projection function in the ProjectionCAD package. Both take all the discriminants and cross resultants of the input polynomials.  The differences arise when taking coefficients from a polynomial in the sub-function Algorithm \ref{alg:2}.  Lazard takes only the leading and trailing coefficients as opposed to all coefficients. Hence \(\textit{ProjL}(\mathcal{F}) \subset \textit{ProjM}(\mathcal{F}) \). Algorithm \ref{alg:1} iterates to perform the maximum chain or projections and produce the full set of projection polynomials.

\par
We will assume familiarity with the details of projection and lifting for CAD, so that we can focus more on the new contributions. Further details can be found in \cite{jirstrand_1995} or \cite{EWBD14} for the implementation at hand.

\textbf{Notation} We use \textbf{wrt} in place of with respect to. 
\par 
\textbf{Definition \stepcounter{Count}\arabic{Count}} Set \texttt{n\_elements}($Polys$) to be the number of elements in the set of polynomials $Polys$. 
\par 
\textbf{Definition \stepcounter{Count}\arabic{Count}} Set \texttt{degree}$(F,x)$ to be the degree of the polynomial $F$, wrt to variable x. 
\par 
\textbf{Definition \stepcounter{Count}\arabic{Count}} Set \texttt{RR}$(x)$ to be the function which returns a set of unique real roots of polynomial x. 
\par  
\textbf{Definition \stepcounter{Count}\arabic{Count}} Set \texttt{R}$(x)$ to be the function which reduces polynomial x to its square free unique factors, with constant multiples removed. 
\par 
\textbf{Notation} The three core algorithms implemented for the classical Lazard projection were: \texttt{ProjectionPolys}, \texttt{Projection} and \texttt{CoefficientSet}. \\
\texttt{ProjectionPolys} is effectively \(\pi^{*}\), \texttt{Projection} is \(\pi_{n_2}\), where $n_2$ is a parameter specified in the input and \texttt{CoefficientSet} which collects the coefficients required for the $n_2$'th Lazard projection. 

\subsection{Worked example}

Before introducing the reader to the generalised algorithm, we will first go through, step by step, a bi-variate case of projection. We will later use this example to build upon for the incremental instance. 
We begin by calculating the projection from the following polynomial system (\ref{eq:f1}):
\begin{equation}
\label{eq:f1}
\textbf{F}_1=\{\underbrace{x_1^2+x_2^2-1}_{f_1},\underbrace{x_1^3-x_2^2}_{f_2}\}.
\end{equation}
Here, the first polynomial is denoted $f_1$ and second denoted $f_2$. We first go through and calculate the univariate polynomials produced by projection and the corresponding real roots. Later, we will use this projection set to lift and build a CAD. We have left out the explicit output polynomials to save space but they can be easily calculated by most computer algebra systems.

\begin{equation}
  \begin{array}{ll}
    r1=\texttt{Resultant}(f_1,f_2,x_2), \ \ \ & \texttt{RR}(r1)=\{\alpha_1\} \approx \{0.7549\}\\ 
    d1=\texttt{Discriminant}(f_1,x_2), \ \ \ &\texttt{RR}(d1)=\{0\}\\
    d2=\texttt{Discriminant}(f_2,x_2), \ \ \ &\texttt{RR}(d2)=\{-1,1\}\\
    l1=\texttt{Leading\ Coefficient}(f_1,x_2), \ \ \ &\texttt{RR}(l1)=\{\}\\
    l2=\texttt{Leading\ Coefficient}(f_2,x_2), \ \ \ &\texttt{RR}(l2)=\{\}\\
    t1=\texttt{Trailing\ Coefficient}(f_2,x_2), \ \ \ &\texttt{RR}(t1)=\{-1,1\}\\
    t2=\texttt{Trailing\ Coefficient}(f_2,x_2), \ \ \ &\texttt{RR}(t2)=\{0\}\\ 
    \end{array}
\end{equation}
So the complete set of isolated real roots is the union: 
$
\{-1,0,\alpha_1 \approx 0.7549,1\} \\
$

In the presentation above we assigned an irrational root to a symbol and gave a decimal approximation, but the implementation would contain the exact root as an algebraic number.  

Figures 3 and 4 plots the graphs of these functions along with the real roots isolated.  We see they correspond to geometrically relevant features.

\begin{figure}[H]
  \centering
  \begin{minipage}[b]{0.48\textwidth}
    \includegraphics[width=\textwidth]{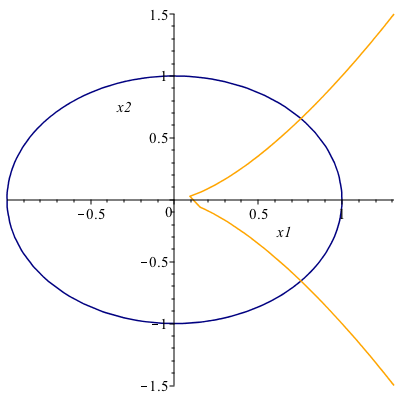}
    \caption{
    The blue curve is $f_1$ and the orange curve is $f_2$.}
  \end{minipage}
  \hfill
  \begin{minipage}[b]{0.48\textwidth}
    \includegraphics[width=\textwidth]{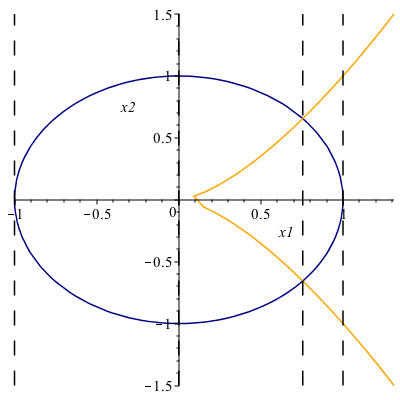}
    \caption{Dotted lines show the projection roots.}
  \end{minipage}
\end{figure}

\subsection{Incremental Lazard projection}

The most appealing approach for creating InCAD was to adapt the Lazard Projection algorithms only to calculate the new projection polynomials. To do this Algorithm \ref{alg:1} and Algorithm \ref{alg:2} had to be modified.
The aim was to directly take the output from the previous call of \texttt{Projection}, together with the incremental polynomial and receive the incremented projection set. We outline the four required adjustments below, how we implemented these adjustments, as well as indicate their active state within the algorithm by highlighting the changes to the algorithms in blue in Algorithms \ref{alg:5} and \ref{alg:4} (\texttt{ProjectionAdd} and \texttt{ProjectionPolysAdd}).


\texttt{ProjectionPolysAdd} is effectively \(\pi^{*}\) for the incremental case, \texttt{ProjectionAdd} is \(\pi_{n_2}\), where $n_2$ is a parameter specified in the input. Note that \texttt{CoefficientSet} itself was unchanged, only the argument passed to it was different. The new argument is now polynomials exclusively from the new polynomial set. 

\subsubsection{Worked example}

In a similar fashion to Section $2.2$, we will guide the reader through a basic incremental example. We will start by incrementing the polynomial system $\textbf{F}_1$, with the polynomial forming a linear line, $f_3=x_2-x_1$. This forms the new system (\ref{eq:4}).
\begin{equation}
\label{eq:4}
\textbf{F}_2=\{\underbrace{x_1^2+x_2^2-1,x_1^3-x_2^2}_{\textbf{F}_1},\underbrace{x_2-x_1}_{f_3}\} 
\end{equation}
The additional projection polynomials and their real roots are:
\begin{equation}
  \begin{array}{ll}
    r2=\texttt{Resultant}(f_1,f_3,x_2), \ \ \  &\texttt{RR}(r2)=\{\pm \alpha_2= \pm \frac{1}{\sqrt[]{2}} \approx \pm 0.7071\} \\
    r3=\texttt{Resultant}(f_2,f_3,x_2), \ \ \  &\texttt{RR}(r3)=\{-1,0\}\\ 
    d3=\texttt{Discriminant}(f_3,x_2), \ \ \ & \texttt{RR}(d3)=\{\}\\
    l3=\texttt{Leading\ Coefficient}(f_3,x_2), \ \ \ &\texttt{RR}(l3)=\{\}\\
    t3=\texttt{Trailing\ Coefficient}(f_3,x_2), \ \ \ &\texttt{RR}(t3)=\{0\}\\
    \end{array}
\end{equation}
So we have rediscovered two roots seen before ($0$ and $-1$) and two new ones ($\pm \alpha_2$).  The total ordered set of real roots is now
\[
\{-1,-\alpha_2 \approx -0.7071, 0, \alpha_2 \approx 0.7071,\alpha_1 \approx 0.7549 ,1\} \\ 
\]
%

Figures 5 and 6 show that the two new roots correspond to the two new intersections of the straight line with the circle.  The other new intersections happened to coincide with other already identified features.

\begin{figure}[H]
  \centering
  \begin{minipage}[b]{0.48\textwidth}
    \includegraphics[width=\textwidth]{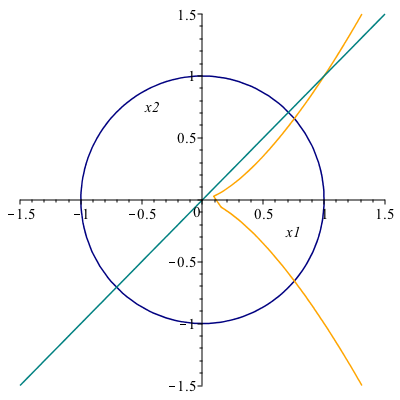}
    \caption{The blue curve is $f_1$, the orange $f_2$ and the teal curve $f_3$.}
  \end{minipage}
  \hfill
  \begin{minipage}[b]{0.48\textwidth}
    \includegraphics[width=\textwidth]{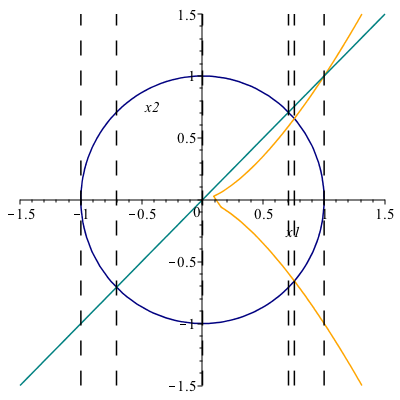}
    \caption{Dotted lines show the projection roots $\in \mathcal{R}$.}
  \end{minipage}
\end{figure}

\subsubsection{Algorithms}

The adjustments required to transfer to an incremental Lazard projection system are detailed as follows: 
\begin{enumerate} 
\item To output the full table of projection polynomials organised by the main variable and reinput this with incremental calls.  This output would then be used as an input for \texttt{ProjectionPolysAdd}. 
\item The process and pass the new polynomials into \texttt{ProjectionAdd}.
\item To create a storage system of new polynomials which we could then apply the minimal number of operations required to obtain the correct Psetnew[i], for each index of i. We introduced a new table called Pset[3] storing the new polynomials created at each iteration of Algorithm \ref{alg:1}. 
\end{enumerate}

\subsection{Incremental projection results}

Notice there were seven elements of the projection set to calculate in the original projection system \texttt{ProjL($F_1$)}. If we were to recompute the incremental system \texttt{ProjL($F_2)$} using the standard method, we would have to compute twelve elements of the projection (the original 7 plus 5 more).  This would be more than double the workload of computing just those 5.   

We split the testing into two experiments, one where we tested the projection algorithm on a test set of 80 examples, the other of 60 more straightforward cases. The first test set is of 80 pairs of tri-variate polynomial couples, each with four terms. The second test set contained 60 pairs of bi-variate polynomials, each with three variates. A pair consists of polynomial a and polynomial b, where b is the polynomial added incrementally. These pairs were created using the random polynomial generating function in Maple. The testing code is openly available\footnote{GitHub repository which includes analysis and worksheet - \newline $\href{url}{https://github.com/acr42/InCAD.git}$}.

An example from the tri-variate, four-term test (with chosen variable ordering $[z, y, x]$):
\begin{equation}
\label{eq:tri-var}
    [-79x^2y^3+97y^2z^2+51x^2, 71x^3y^2+22xz-47y^2]
\end{equation}
An example from the bi-variate, three-term test (with variable ordering $[z, y]$):
\begin{equation}
\label{eq:bi-var}
    [66z^5-40y^4+75yz^2+83z, -33y^2z^3+72y^3z+84yz^2-4z^2]
\end{equation}
The tests were controlled by a bash script rather than code within a single Maple session to avoid any caching which could bias the results.

\subsubsection{Experiment: Tri-variate}

When looking at the cases which were faster, they were on average faster by $30\%$, and $39\%$ slower in the other cases. The net speed difference had a mean of $28\%$ increase in performance with the incremental projection.

\begin{center}
\begin{tabular}{@{}*4l@{}}
  \toprule[1.5pt]
  \multicolumn{1}{c}{\head{Projection}} &
    \multicolumn{2}{c}{\head{Results}} \\ 
  \head{} & \head{Classical} & \head{Incremental} & \head{}\\
  \cmidrule(l){1-1}\cmidrule(r){2-4}
  \verb|Variance| & \rmfamily 0.002743s & \rmfamily  0.002205s & \rmfamily  \textcolor{green}{\textbf{24.39\%}} Smaller\\
  \verb|Mean| & \rmfamily 0.06739s & \rmfamily  0.04809s & \rmfamily  \textcolor{green}{\textbf{28.64\%}} Faster \\
  \verb|Lower Quartile| & \rmfamily 0.02475s & \rmfamily  0.013s & \rmfamily  \textcolor{green}{\textbf{47.47\%}} Faster \\
  \verb|Median| & \rmfamily 0.0625s & \rmfamily  0.035s & \rmfamily  \textcolor{green}{\textbf{44.00\%}} Faster\\
  \verb|Upper Quartile| & \rmfamily 0.09425s & \rmfamily  0.07525s & \rmfamily  \textcolor{green}{\textbf{20.16\%}} Faster \\
  \bottomrule[1.5pt]
\end{tabular}
\end{center}

\subsubsection{Experiment: Bi-variate} 
When looking at the cases which were faster, they were on average faster by $55\%$, and $87\%$ slower in the other cases. The net speed difference had a mean of $16\%$ increase in performance with the incremental projection.
\par 

\begin{center}
\begin{tabular}{@{}*4l@{}}
  \toprule[1.5pt]
  \multicolumn{1}{c}{\head{Projection}} &
    \multicolumn{2}{c}{\head{Results}} \\ 
  \head{} & \head{Classical} & \head{Incremental} & \head{}\\
  \cmidrule(l){1-1}\cmidrule(r){2-4}
  \verb|Variance| & \rmfamily 0.0004660s & \rmfamily  0.0006425s & \rmfamily  \textcolor{red}{\textbf{27.46\%}} Larger\\
  \verb|Mean| & \rmfamily 0.03743s & \rmfamily  0.0315s & \rmfamily  \textcolor{green}{\textbf{15.85\%}} Faster \\
  \verb|Lower Quartile| & \rmfamily 0.024s & \rmfamily  0.008s & \rmfamily  \textcolor{green}{\textbf{66.66\%}} Faster \\
  \verb|Median| & \rmfamily 0.0285s & \rmfamily  0.015s & \rmfamily  \textcolor{green}{\textbf{47.39\%}} Faster\\
  \verb|Upper Quartile| & \rmfamily 0.03675s & \rmfamily  0.05525s & \rmfamily  \textcolor{red}{\textbf{50.34\%}} Slower \\
  \bottomrule[1.5pt]
\end{tabular}
\end{center}

\begin{algorithm}
\caption{ProjectionPolys}
\label{alg:1}
{\fontsize{10}{15}\selectfont
\begin{algorithmic}[1]
\STATE \textbf{Input:} A system of polynomials $polyset=\{f_1,\ldots,f_m\} \in \mathbb{R}[x_n,\ldots,x_1]$, $var=[x_n,\ldots,x_1]$
\STATE \textbf{Output:} A list of projection polynomials $\in \mathbb{R}[x_{n-1}, \dots, x_{1}] $
\STATE \textbf{Procedure} \texttt{Lazard ProjectionPolys}
\COMMENT{"Compute the full chain of projections"}
\STATE $dim \leftarrow$ Number of elements in $var$
\STATE $pset[0] = table()    $ 
\STATE $pset[0] \leftarrow$ Primitive set from $polyset$, wrt variable $x_n$ 
\STATE $pset[0] \leftarrow$ Square free basis set from $pset[0]$, wrt variable $x_n$ 
\STATE $pset[0] \leftarrow$ Set of factors from $pset[0]$, wrt variable $x_n$ 
\STATE $cont \leftarrow$ Content Set from $polyset$, wrt variable $x_n$ 
\FOR{\texttt{i from 1 to dim-1}}
\STATE $out \leftarrow$ \texttt{Projection}$(pset[i-1], x_{n-i+1}, [x_{n-i},\ldots,x_1])$ 
\STATE $pset[i] \leftarrow (out \ \cup \ cont)$
\STATE $cont \leftarrow$ Content set from $pset[i]$, wrt $x_{n-i}$ 
\STATE $pset[i] \leftarrow$ Primitive set from $pset[i]$, wrt variable $x_{n-i}$
\STATE $pset[i] \leftarrow$ Square free basis set from $pset[i]$, wrt variable $x_{n-i}$ 
\STATE $pset[i] \leftarrow$ Set factors from $pset[i]$, wrt variable $x_{n-i}$ 
\ENDFOR

\STATE $ret \leftarrow pset[dim-1]$ 
\STATE $ret    \leftarrow$ Remove constant multiples from $ret$
\STATE \textbf{return} $ret$
\end{algorithmic}
}
\end{algorithm}

\begin{algorithm}
\caption{Projection}
\label{alg:2}
{\fontsize{10}{15}\selectfont
\begin{algorithmic}[1]
\STATE \textbf{Input:} A set of polynomials $polyset=\{f_1,\ldots,f_m\} \in \mathbb{R}[x_n,\ldots,x_1]$, $var=x_n$ and $lvars=[x_{n-1},\ldots,x_1]$
\STATE \textbf{Output:} A set of projection polynomials $\mathbb{P}=\{p_1,\ldots,p_q\} \in \mathbb{R}[x_{n-1},\ldots,x_{1}] $
\STATE \textbf{Procedure} \texttt{Lazard Projection}
\COMMENT{"Lazard projection operator, \(\pi_{n-1}\)."}
\STATE $Polys \leftarrow$ Primitive set from $polyset$, wrt variable $x_n$ 
\STATE $Cont \leftarrow$ Content set from $polyset$, wrt variable $x_n$ 
\STATE $Polys \leftarrow$ Square free basis from $Polys$, wrt $x_n$ 
\STATE $Pset1 = table()$
\FOR{\texttt{i from 1 to n\_elements(Polys)}}
        \STATE $Pol\leftarrow Polys[i]$
        \STATE $clist \leftarrow$ Lazard coefficient set from $Pol$, wrt variable $x_n$
        \STATE $temp \leftarrow$  Discriminant set from $Pol$, wrt variable $x_n$ 
        \STATE $temp \leftarrow$  Remove constant multiples from $temp$ 
        \STATE $Pset1[i] \leftarrow$ Concatenate together $temp$ and $clist$ \COMMENT{Need to make sure you deal with the case of temp being empty here}
\ENDFOR
\STATE $Pset2 = table()$
\FOR{\texttt{i from 1 to n\_elements(Polys)}}
\FOR{\texttt{j from i+1 to n\_elements(Polys)}}
    \STATE     $Pset2[i,j] \leftarrow$ Resultant of the two polynomials $Polys[i]$ \& $Polys[j]$, wrt variable $var$
    \STATE     $Pset2[i,j] \leftarrow $ Remove constant multiples from $Pset2[i,j]$
\ENDFOR
\ENDFOR
\STATE $Pset \leftarrow cont,Pset1,Pset2$  \COMMENT {Note that $Pset \in \mathbb{R}[x_{n-1},\ldots,x_{1}]$}
\STATE $Pset \leftarrow$ Remove constants from $Pset$ 
\STATE \textbf{return} $Pset$
\end{algorithmic}
}
\end{algorithm}

\begin{algorithm}
\caption{ProjectionPolysAdd}
\label{alg:5}
{\fontsize{10}{15}\selectfont
\begin{algorithmic}[1]
\STATE \textbf{Input:} A system of polynomials $polyset=\{f_1,..f_m\} \in \mathbb{R}[x_n,..,x_1]$, \textcolor{blue}{$prevprojset \in \mathbb{R}[x_n,\ldots,x_{1}]$, $newpolset\in \mathbb{R}[x_{n},\ldots,x_1]$} and $vars=[x_n,..,x_1]$ \textcolor{blue}{(Adjustment 1)}
\STATE \textbf{Output:} A list of projection polynomials $\in \mathbb{R}[x_{n-1},..,x_{1}] $
\STATE \textbf{Procedure} \texttt{Lazard ProjectionPolysAdd}
\COMMENT{"Compute the full chain of projections, \(\pi*\)"}
\STATE $dim \leftarrow$ Number of elements in $vars$ $+1$
\STATE $pset[0] = table()$ 
\STATE $pset[0] \leftarrow$ Primitive set from \textcolor{blue}{$newpolset$}, wrt variable $x_n$ \textcolor{blue}{(Adjustment 2)}
\STATE $pset[0] \leftarrow$ Square free basis set from $pset[0]$, wrt variable $x_n$ 
\STATE $pset[0] \leftarrow$ Set of factors from $pset[0]$, wrt variable $x_n$ 
\STATE $cont \leftarrow$ Set of contents of \textcolor{blue}{$newpolset$}, wrt $x_n$ \textcolor{blue}{(Adjustment 2)}
\FOR{\texttt{i from 1 to dim-1}}
\STATE $out \leftarrow$ \texttt{ProjectionAdd}$(\textcolor{blue}{prevprojset[i-1]},pset[i-1], x_{n-i+1},\ldots,x_1)$ \\ \textcolor{blue}{(Adjustment 2)}
\STATE $pset[i] \leftarrow (out \ \cup \ cont)$
\STATE $cont \leftarrow$ Content set from $pset[i]$, wrt variable x$_{n-i}$ 
\STATE $pset[i] \leftarrow$ Prime set from $pset[i]$, wrt variable $x_{n-i}$
\STATE $pset[i] \leftarrow$ Square free basis from set $pset[i]$, wrt variable $x_{n-i}$ 
\STATE $pset[i] \leftarrow$ Set factors from $pset[i]$, wrt variable $x_{n-i}$ 
\ENDFOR

\STATE $pset[dim-1]    \leftarrow$ Remove constant multiples from $pset[[dim-1]$
\STATE $ret \leftarrow pset[[dim-1]$ 

\STATE \textbf{return} \textcolor{blue}{{$pset$} (Adjustment 1)}
\end{algorithmic}
}
\end{algorithm}

\begin{algorithm}
\caption{ProjectionAdd}
\label{alg:4}
{\fontsize{10}{15}\selectfont
\begin{algorithmic}[1]
\STATE \textbf{Input:} Polynomial set $newpprojset=\{f_1,..f_m\} \in \mathbb{R}[x_n,\ldots,x_1]$, \textcolor{blue}{$oldprojset \in \mathbb{R}[x_{n-1},\ldots,x_{1}]$}, and variable ordering 
\textcolor{blue}{(Adjustment 2)}
\STATE \textbf{Output:} Set of polynomials $Pset=\{p_1,\ldots,p_q\} \in \mathbb{R}[x_{n-1},\ldots,x_{1}] $
\STATE \textbf{Procedure} \texttt{Lazard ProjectionAdd}
\COMMENT{The  incremental version of the Lazard projection operator, \(\pi_{n-1}\).}
\STATE $Polys \leftarrow$ Primitive set from \textcolor{blue}{$newpprojset$}, wrt variable $x_n$ \textcolor{blue}{(Adjustment 2)}
\STATE $Cont \leftarrow$ Content set from \textcolor{blue}{$newpprojset$}, wrt variable $x_n$ \textcolor{blue}{(Adjustment 2)}
\STATE $Polys \leftarrow$ Square free basis set from $Polys$, wrt variable $x_n$ 
\STATE $Pset1 = table():$
\FOR{\texttt{i from 1 to number\_elements(Polys)}}
        \STATE $Pol\leftarrow Polys[i] $
        \STATE $clist \leftarrow $ Lazard coefficient set from $Pol$, wrt to $x_n$
        \STATE $temp \leftarrow$ Discriminant set from $Pol$, wrt $x_n$ 
        \STATE $temp \leftarrow$ Remove constant multiples from $temp$ 
        \STATE $Pset1[i] \leftarrow$ concat $temp$ \& $clist$ 
\ENDFOR
\STATE $Pset2 = table():$
\FOR{\texttt{i from 1 to n\_elements(Polys)}}
\FOR{\texttt{j from i+1 to n\_elements(Polys)}}
    \STATE     $Pset2[i,j] \leftarrow$ Resultant of $Polys[i]$ and $Polys[j]$, wrt to variable $var$
    \STATE     $Pset2[i,j] \leftarrow$ Remove constant multiples from $Pset2[i,j]$
\ENDFOR
\ENDFOR \textcolor{blue}{
\STATE $oldset \leftarrow oldprojset $ 
\STATE $Pset3 = table():$
\FOR{\texttt{i from 1 to n\_elements(Polys)}}
\FOR{\texttt{j from 1 to n\_elements(oldset)}}
    \STATE     $Pset3[i,j] \leftarrow $ Resultant of $Polys[i]$ and $oldset[j]$, wrt to variable $var$
    \STATE     $Pset3[i,j] \leftarrow $ Remove constant multiples from $Pset3[i,j]$
\ENDFOR
\ENDFOR \COMMENT{Adjustment 3}}
\STATE $Pset \leftarrow$ concat $(cont,Pset1,Pset2,\textcolor{blue}{Pset3})$  \COMMENT{$Pset \in \mathbb{R}[x_n,\ldots,x_{1}]$}
\STATE $Pset \leftarrow$ Remove constant multiples from $Pset$ 
\STATE \textbf{return} $Pset$
\end{algorithmic}
}
\end{algorithm}

\newpage

\section{Lifting}

\subsection{Lifting after Lazard projection}

We had to make changes to the lifting code not just to allow for incrementality but also changes required by the use of Lazard projection as set out in \cite{MPP17}.  This requires the use of Lazard evaluation of polynomials (Algorithm \ref{alg:7}).  To simplify we restricted our implementation to the open case.

\textbf{Definition \stepcounter{Count}\arabic{Count}} An \textbf{Open-CAD} is one produced by lifting over open intervals only.  Thus it is not a full decomposition of $\R^n$ as the boundaries of the full dimensional cells are missing.

The advantage is that when building an Open-CAD we need never extend over irrational sample points, avoiding costly algebraic number calculations, but still getting a good understanding of the solution set.

We will now discuss a method used for incremental lifting, which can be thought of graphically, as a form of acyclic tree merge, later displayed visually. It will first make more sense if we explicitly show the reader the mathematical structure that allows CAD to be represented as a tree. Keeping the structure shown here in mind, will in hope make most of the algorithmic decisions, implemented during incremental lifting, much clearer. 

We have two core algorithms for the Lifting. First the initial \texttt{LiftSetup} set-up algorithm (Algorithm \ref{alg:6}), and then \texttt{Lift} (Algorithm \ref{alg:8}) which iterates over the variable ordering until completion. There are also a number of other less complicated vital sub algorithms used for cell formatting/sorting/lifting purposes such \texttt{NewCadCells}, which creates new cells when lifting to a set of given roots, \texttt{GenSamplePoints} for generating sample points, \texttt{SubsFormat} which formats the sample points for substitution.

\subsection{Worked example and Lazard evaluation}

\textbf{Definition \stepcounter{Count}\arabic{Count}} Let \texttt{divide}$(x,d)$ to be the function returns boolean value true if $x$ is divisible by polynomial $d$, otherwise false. 

\textbf{Definition \stepcounter{Count}\arabic{Count}} Let \texttt{SDivide}$(x,d,v)$ to be the function will completely divide polynomial $d$; which is for example $\in \mathcal{R}[x_1]$, out of polynomial $x \in \mathcal{R}[x_1,x_2]$, until the output is no longer divisible by $d$. in which  case \texttt{SDivide} will then take this polynomial and substitutes the value $v$ into all instances of the variable $x_1$, returning a square free univariate polynomial in $x_2$. 

We will guide the reader through a primary Lazard lift example, using Lazard evaluation (Algorithm \ref{alg:7}).  The key difference to traditional CAD lifting is that we must calculate the Lazard evaluation of each polynomial at each sample point.  This avoids the well-orientedness issues of the McCallum projection operators (where information is lost due to nullification at a point). 

\begin{algorithm}[h]
\caption{Lazard evaluation}
\label{alg:7}
{\fontsize{10}{15}\selectfont
\begin{algorithmic}[1]
\STATE \textbf{Input:} A polynomial from the projection set $f \in \mathbb{R}[x_1,\ldots,x_d]$, and $\texttt{samplepoints}=[r_1,\ldots,r_{d-1}] \in \mathbb{R}^{d-1}$.
\STATE \textbf{Output:} List of roots. 
\STATE \textbf{Procedure} \texttt{Lazard\_evaluation}:
\STATE Set $\texttt{roots}$ to be an empty list 
\FOR {\texttt{$j$ from $1$ to $d-1$}}
\STATE Break if cell is zero dimensional.
\FOR {\texttt{$i$ from $1$ to degree\( (f,x_j) \)}}
\IF {\texttt{f is divisible by $x_j-r_j$}}
\STATE $f \leftarrow f/(x_j-r_j)$ 
\ELSE 
\STATE \textbf{break} innermost for loop: 
\ENDIF
\ENDFOR 
\ENDFOR 
\STATE $f \leftarrow$ Substitute the $\texttt{samplepoints}$ into $f$ 
\STATE \COMMENT{$f \in \mathbb{R}[x_n,\ldots,x_d] $}
\STATE $roots \leftarrow$ Real roots of $f$
\STATE $roots \leftarrow$ Remove duplicates and sort $roots$ in ascending order
\STATE \textbf{return} roots:
\end{algorithmic}
}
\end{algorithm}

We will start by lifting the projection polynomial system defined previously \texttt{ProjL}($\textbf{F}_1$), described below. In our implementation, we only performed the lift over open intervals, as we followed the Open-CAD method. However, for this example, we will lift all cells. 

\begin{equation}
    \texttt{ProjL}(\textbf{F}_1)=\{x_1+1,x_1,x_1-\alpha_1 ,x_1-1\}
\end{equation}

First, we must generate our sample points for the decomposition of the real line with respect to the roots $\{-1,0,\alpha_1,1\}$.
Thus, we need a sample value from each of the following cells:
\begin{equation}
\label{ineq:1}
  \begin{array}{lll}
     a1=\{ x_1<-1\}, &a2=\{x_1=-1\}, &a3=\{-1<x_1<0\},\\
    a4=\{x_1=0\},&a5=\{0<x_1<\alpha_1\}, &a6=\{x_1=\alpha_1\},\\
     a7=\{\alpha_1<x_1<1\}, &a8=\{x_1=1\},&a9= \{1<x_1\} 
    \end{array}
\end{equation}

Note: $ai$ denotes the sorted $i$'th cell in the decomposition of $x_1$-space.

Our nine chosen sample points are:
\begin{equation}
  \begin{array}{l}
    \texttt{SamplePoints}(\texttt{ProjL}(\textbf{F}_1))=\{-2,-1,-0.5,0,0.5,\alpha_1 \approx 0.7549,0.9,1,2 \} \\ 
    \end{array}
\end{equation}
Now we lift over each cell at the designated sample point by isolating real roots of the corresponding univariate polynomials.  Below, $p_{i,j}$ denotes the polynomial acquired after applying the Lazard evaluation method to the $i$'th sample point, on the $j$'th polynomial from $F_1$ 
    
\begin{equation*}
  \begin{array}{l}
    \textnormal{Performing lift on the first sample point of } $$-2:$$ \\
    p_{1,1}=\texttt{SDivide}(f_1,(x_1+2),-2), \ \ \ \texttt{RR}(p_{1,1})=\{\} \\
    p_{1,2}=\texttt{SDivide}(f_2,(x_1+2),-2), \ \ \ \texttt{RR}(p_{1,2})=\{\} \\
    \\
    \textnormal{Performing lift on the next sample point } $$-1:$$ \\
    p_{2,1}=\texttt{SDivide}(f_1,(x_1+1),-1), \ \ \ \texttt{RR}(p_{2,1})=\{0\} \\
    p_{2,2}=\texttt{SDivide}(f_2,(x_1+1),-1), \ \ \ \texttt{RR}(p_{2,2})=\{\} \\
    \\
    \textnormal{Performing lift on the next sample point } $$-0.5:$$ \\
    p_{3,1}=\texttt{SDivide}(f_1,(x_1+0.5),-0.5), \ \ \ \texttt{RR}(p_{3,1})=\{-\beta_0 \approx -0.8660,\beta_0 \approx 0.8660\} \\
    p_{3,2}=\texttt{SDivide}(f_2,(x_1+0.5),-0.5), \ \ \ \texttt{RR}(p_{3,2})=\{\} \\
    \\
    \textnormal{Performing lift on the next sample point } $$0:$$ \\
    p_{4,1}=\texttt{SDivide}(f_1,(x_1),0), \ \ \ \texttt{RR}(p_{4,1})=\{-1,1\} \\
    p_{4,2}=\texttt{SDivide}(f_2,(x_1),0), \ \ \ \texttt{RR}(p_{4,2})=\{\} \\
    \end{array}
\end{equation*}
\begin{equation*}
  \begin{array}{l}
    \textnormal{Performing lift on the next sample point } $$0.5:$$ \\
       p_{5,1}=\texttt{SDivide}(f_1,(x_1-0.5),0.5), \ \ \ \texttt{RR}(p_{5,1})=\{-\beta_1 \approx -0.8660,\beta_1 \approx 0.8660\} \\
    p_{5,2}=\texttt{SDivide}(f_2,(x_1-0.5),0.5), \ \ \ \texttt{RR}(p_{5,2})=\{-\beta_2 \approx -0.3536,\beta_2 \approx 0.3536\} \\ 
    \\ 
    \textnormal{Performing lift on the next sample point } \alpha_1: \\
    p_{6,1}=\texttt{SDivide}(f_1,(x_1-0.5),0.5), \ \ \ \texttt{RR}(p_{5,1})=\{-\beta_3 \approx -0.6559,\beta_3 \approx 0.6559\} \\
    p_{6,2}=\texttt{SDivide}(f_2,(x_1-0.5),0.5), \ \ \ \texttt{RR}(p_{5,2})=\{-\beta_4 \approx -0.6559,\beta_4 \approx 0.6559\} \\
    \\
    \textnormal{Performing lift on the next sample point } $$0.9:$$ \\
    p_{7,1}=\texttt{SDivide}(f_1,(x_1-0.9),0.9), \ \ \ \texttt{RR}(p_{7,1})=\{-\beta_5 \approx -0.4359,\beta_5 \approx 0.4359\} \\
    p_{7,2}=\texttt{SDivide}(f_2,(x_1-0.9),0.9), \ \ \ \texttt{RR}(p_{7,2})=\{-\beta_6 \approx -0.8538,\beta_6 \approx 0.8538\} \\
    \\
    \textnormal{Performing lift on the next sample point } $$1:$$ \\
    p_{8,1}=\texttt{SDivide}(f_1,(x_1-1),1), \ \ \ \texttt{RR}(p_{8,1})=\{0\} \\
    p_{8,2}=\texttt{SDivide}(f_2,(x_1-1),1), \ \ \ \texttt{RR}(p_{8,2})=\{-1,1\} \\
    \\
    \textnormal{Performing lift on the next sample point } $$2:$$ \\
    p_{9,1}=\texttt{SDivide}(f_1,(x_1-2),2), \ \ \ \texttt{RR}(p_{9,1})=\{\} \\
    p_{9,2}=\texttt{SDivide}(f_2,(x_1-2),2), \ \ \ \texttt{RR}(p_{9,2})=\{-\beta_7 \approx -2.8284,\beta_7 \approx 2.8284\} 
    \end{array}
\end{equation*}

\begin{figure}[H]
  \centering
  \begin{minipage}[b]{\textwidth}
    \includegraphics[width=\textwidth]{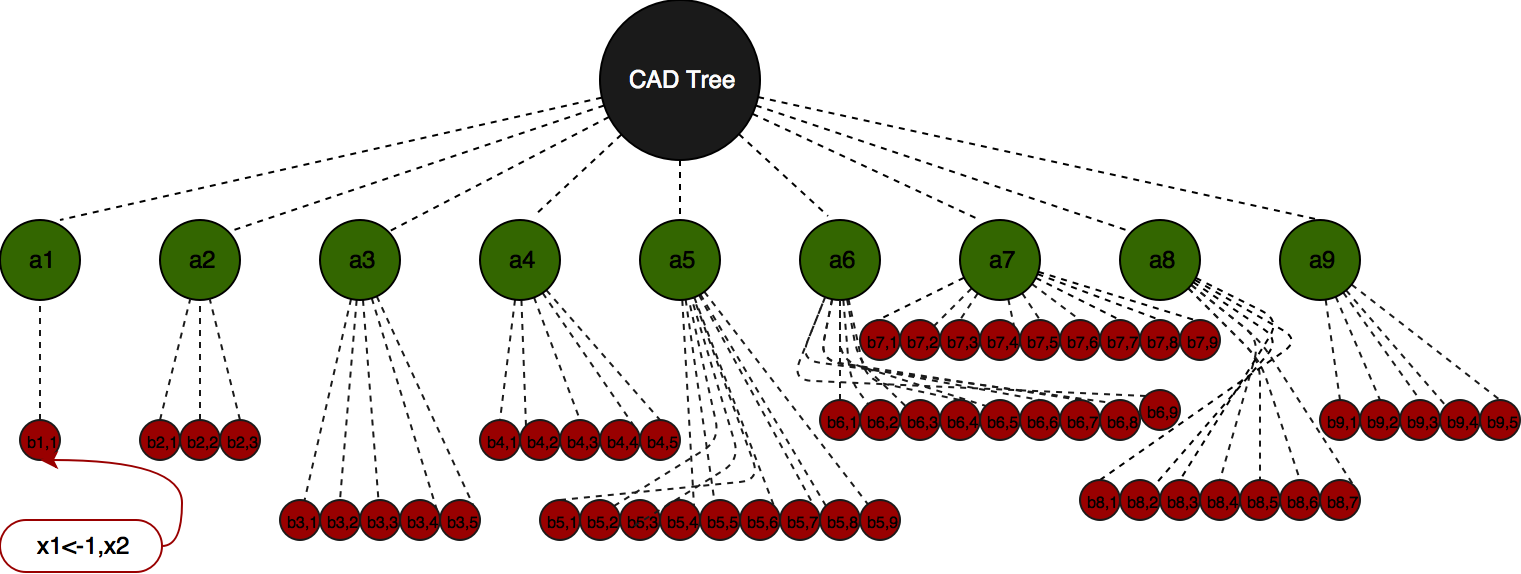}
    \caption{CAD tree of $F_1$. The green nodes represent cells in the first dimension and red nodes represent cells in the second dimension.}
    \label{tree:1}
  \end{minipage} 
\end{figure}

We now have enough information to generate our CAD cells in $\R^2$. To describe our full CAD system more concisely, we will now define a new set of inequalities (\ref{ineq:inc}) we have acquired from the lift stage into $x_2$. For simplicity, we will display the CAD's created throughout this article using a graphical tree representation Fig (\ref{tree:1}).  Here $bi,j$ denotes the sorted $j$'th inequality we have decomposed the $ai$'th inequality into, and so $[i,j]$ corresponds to the classical CAD cell index.

The new cell definitions are as given below. 
\begin{align}
\label{ineq:inc}
&\textrm{Inequalities}(\textbf{F}_1) = \\
  &\begin{array}{l}
    b_{1,1}=\{x_2\}, \,
        b_{2,1}=\{x_2<0\},b_{2,2}=\{x_2=0\}, \,
        b_{2,3}=\{0<x_2\},\\
        b_{3,1}=\{x_2<-\beta_0\},b_{3,2}=\{x_2=-\beta_0\},b_{3,3}=\{-\beta_0<x_2<\beta_0\},\\
        b_{3,4}=\{x_2=\beta_0\},b_{3,5}=\{\beta_0<x_2 \}, \,
        b_{4,1}=\{x_2<-1\},b_{4,2}=\{x_2=-1\},\\
        b_{4,3}=\{-1<x_2<1\},b_{4,4}=\{x_2=1\},b_{4,5}=\{1<x_2\},\\
         b_{5,1}=\{x_2<-\beta_1\},b_{5,2}=\{x_2=-\beta_1\},b_{5,3}=\{-\beta_1<x_2<-\beta_2\},\\
         b_{5,4}=\{x_2=-\beta_2\},b_{5,5}=\{-\beta_2<x_2<\beta_2 \},b_{5,6}=\{x_2=\beta_2 \},\\
         b_{5,7}=\{\beta_2<x_2<\beta_1 \},b_{5,8}=\{x_2=\beta_1 \},b_{5,9}=\{\beta_1<x_2\},\\
         b_{6,1}=\{x_2<-\beta_3\},b_{6,2}=\{x_2=-\beta_3\},b_{6,3}=\{-\beta_3<x_2<-\beta_4\},\\
         b_{6,4}=\{x_2=-\beta_4\},b_{6,5}=\{-\beta_4<x_2<\beta_4 \},b_{6,6}=\{x_2=\beta_4 \},\\
         b_{6,7}=\{\beta_4<x_2<\beta_3 \},b_{6,8}=\{x_2=\beta_3 \},b_{6,9}=\{\beta_3<x_2\},\\
        b_{7,1}=\{x_2<-\beta_6\},b_{7,2}=\{x_2=-\beta_6\},b_{7,3}=\{-\beta_6<x_2<-\beta_5\},\\
        b_{7,4}=\{x_2=-\beta_5\},b_{7,5}=\{-\beta_5<x_2<\beta_5 \},b_{7,6}=\{x_2=\beta_5 \},\\
        b_{7,7}=\{\beta_5<x_2<\beta_6\},b_{7,8}=\{x_2=\beta_6 \},b_{7,9}=\{\beta_6<x_2\},\\
        b_{8,1}=\{x_2<-1\},b_{8,2}=\{x_2=-1\},b_{8,3}=\{-1<x_2<0\},\\
        b_{8,4}=\{x_2=-0\},b_{8,5}=\{0<x_2<1 \},b_{8,6}=\{x_2=1 \},b_{8,7}=\{1<x_2\},\\
        b_{9,1}=\{x_2<-\beta_7\},b_{9,2}=\{x_2=-\beta_7\},b_{9,3}=\{-\beta_7<x_2<\beta_7\},\\
        b_{9,4}=\{x_2=\beta_7\},b_{9,5}=\{\beta_7<x_2
    \}.
    \end{array} \nonumber
\end{align}

\newpage

\subsection{Incremental Lazard lifting}

The two novel algorithms created for incremental lifting were the \newline \texttt{LiftSetupAdd} (Algorithm \ref{alg:10}); the incremental version of \texttt{LiftSetup} (Algorithm \ref{alg:6}), and \texttt{LiftAdd} (Algorithm \ref{alg:liftadd}); the incremental version of \texttt{Lift} (Algorithm \ref{alg:8}). 
\par 

The general concept of how we solved this stage of the problem was to think of it as solving a graph (tree) attachment/detachment problem. If you think of the old CAD as having a tree structure which we save, where nodes are cells, and branches link a cell to its parent (cell it projects onto) or child (decomposition in a cylinder above) cells. At each depth of the CAD/tree, say depth $p$, are all the cells within $\mathbb{R}^{p}$ before we lifted to $\mathbb{R}^{p+1}$. We first go through a worked example before describing the general process.
\par 


We will now perform an incremental lift on the polynomial system $\textbf{F}_1$, incremented by a new polynomial $f_4=x_1^3+x_2^2$, forming the new system (\ref{eq:f3}). 

\begin{equation}
    \textbf{F}_3=\{\underbrace{x_1^2+x_2^2-1,x_1^3-x_2^2}_{\textbf{F}_1},\underbrace{x_1^3+x_2^2}_{f_4}\} 
    \label{eq:f3}
\end{equation}

The new system is symmetrical about the $y$ axis as you can see in Figure 8. 

\begin{figure}[H]
  \centering
  \begin{minipage}[b]{0.4\textwidth}
    \includegraphics[width=\textwidth]{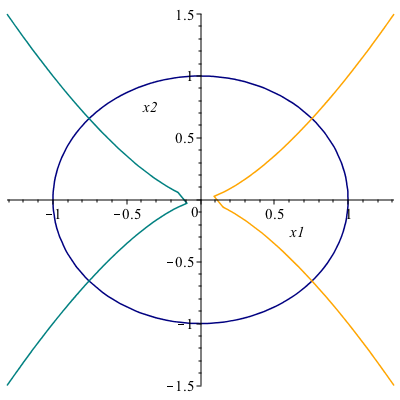}
    \caption{
    The blue curve is $f_1$, the orange $f_2$ and the teal $f_4$.}
  \end{minipage}
  \hfill
  \begin{minipage}[b]{0.4\textwidth}
    \includegraphics[width=\textwidth]{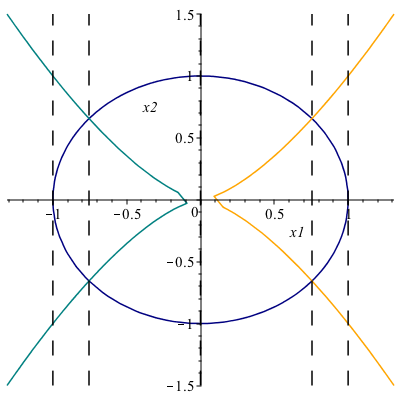}
    \caption{Dotted lines show the projection roots.}
  \end{minipage}
\end{figure}

We will skip the projection steps. However they are contained in detail within the Maple "Worked Examples" worksheet, available on GitHub\footnote{$\href{url}{https://github.com/acr42/InCAD.git}$}.

\begin{equation}
    \texttt{ProjL}(\textbf{F}_3)=\ \{ \underbrace{x_1+1}_{\texttt{ProjL}(\textbf{F}_1)},\underbrace{x_1+\alpha_1}_{new},\underbrace{x_1,x_1-\alpha_1 ,x_1-1}_{\texttt{ProjL}(\textbf{F}_1)}\}
\end{equation}

We will be using calculations from Section 3.2 where we have already performed a full lift on the part of the system, bar the new projection polynomial of $x_1+\alpha_1$.

\textbf{Incremental Lift}

We now have an enlarged set of roots of univariate projection polynomials, 
$\{-1,-\alpha_1,0,\alpha_1,1\}$.  
The new additions below due to adding in $f_4$ are highlighted in blue. 

Decomposition of the real line:
\begin{equation}
  \begin{array}{l}
  \label{ineq:inlift}
     a1=\{ x_1<-1\}, a2=\{x_1=-1\},\textcolor{blue}{ a3=\{-1<x_1<-\alpha_1\}},\\
    \textcolor{blue}{a4=\{x_1=-\alpha_1\}},\textcolor{blue}{a5=\{-\alpha_1<x_1<0\}}, a6=\{x_1=0\},
     a7=\{0<x_1<\alpha_1\}, \\ a8=\{x_1=\alpha_1\},a9= \{\alpha_1<x_1<1\},a10= \{x_1=1\},a11= \{1<x_1\} \\
     \end{array}
\end{equation}
Corresponding sample points: 
\begin{equation}
  \begin{array}{l}
  \label{smple:inclft}
    \texttt{SamplePoints}(ProjL(\textbf{F}_3))=\\
    \{-2,-1,\textcolor{blue}{-0.9},\textcolor{blue}{-\alpha_1},-0.5,0,0.5,\alpha_1 \approx 0.7549,0.9,1,2 \} \\ 
    \end{array}
\end{equation}

We first begin testing whether each fixed sample point has a new root, in the new projection polynomials, due to the incremental polynomial $f_4$. If it does, we will merge the new roots into the old roots list for that cell (in $x_1$). Once we have a list of new and old roots to life over, we then recompute a lift over that cell. If the previous sample point does not have any different roots to that cell,  we keep its structure unchanged. 
\begin{equation}
  \begin{array}{l}
    \textnormal{Performing lift on the first sample point of } $$-2:$$ \\
    p_{1,3}=\texttt{SDivide}(f_3,(x_1+2),-2), \ \ \ \texttt{RR}(p_{1,3})= \textcolor{blue}{ \{\pm \beta_8 \approx \pm 0.8284 \} } \\
    \textnormal{Performing lift on the next sample point } $$-1:$$ \\
    p_{2,3}=\texttt{SDivide}(f_3,(x_1+1),-1), \ \ \ \texttt{RR}(p_{2,3})=\{-1,1\} \\
    \textnormal{Performing lift on the next sample point } $$-0.5:$$ \\
    p_{5,3}=\texttt{SDivide}(f_3,(x_1+0.5),-0.5), \ \ \ \texttt{RR}(p_{5,3})=\textcolor{blue}{\{\pm \beta_9 \approx \pm 0.3536\}} \\
    \textnormal{Performing lift on the next sample point } $$0:$$ \\
    p_{6,3}=\texttt{SDivide}(f_3,(x_1),0), \ \ \ \texttt{RR}(p_{6,3})=\{\} \\
    \textnormal{Performing lift on the next sample point } $$0.5:$$ \\
       p_{7,3}=\texttt{SDivide}(f_3,(x_1-0.5),0.5), \ \ \ \texttt{RR}(p_{7,3})=\{\}
       \\
  \textnormal{Performing lift on the next sample point } \alpha_1: \\
    p_{8,3}=\texttt{SDivide}(f_3,(x_1-0.5),0.5), \ \ \ \texttt{RR}(p_{8,3})=\{\} \\
    \textnormal{Performing lift on the next sample point } $$0.9:$$ \\
    p_{9,3}=\texttt{SDivide}(f_3,(x_1-0.9),0.9), \ \ \ \texttt{RR}(p_{9,3})=\{\} \\
    \textnormal{Performing lift on the next sample point } $$1:$$ \\
    p_{10,3}=\texttt{SDivide}(f_3,(x_1-1),1), \ \ \ \texttt{RR}(p_{10,3})=\{\} \\
    \textnormal{Performing lift on the next sample point } $$2:$$ \\
    p_{11,3}=\texttt{SDivide}(f_3,(x_1-2),2), \ \ \ \texttt{RR}(p_{11,3})=\{\} 
    \end{array}
\end{equation}

We now complete a full lift by considering the new sample points. 
\begin{equation}
  \begin{array}{l}
  \textnormal{Performing lift on the next sample point } $$-0.9,$$ \\
  p_{3,1}=\texttt{SDivide}(f_1,(x_1+0.9),-0.9), \ \ \ \texttt{RR}(p_{3,1})=\textcolor{blue}{\{\pm\beta_{10} \approx \pm 0.4359\} }\\
  p_{3,2}=\texttt{SDivide}(f_2,(x_1+0.9),-0.9), \ \ \ \texttt{RR}(p_{3,2})=\{\} \\
    p_{3,3}=\texttt{SDivide}(f_3,(x_1+0.9),-0.9), \ \ \ \texttt{RR}(p_{3,3})=\textcolor{blue}{\{\pm \beta_{11} \approx \pm 0.8538\} }\\
    \textnormal{Performing lift on the next sample point } -\alpha_1, \\
    p_{4,1}=\texttt{SDivide}(f_1,(x_1+\alpha_1),-\alpha_1), \ \ \ \texttt{RR}(p_{4,1})=\textcolor{blue}{\{\pm \beta_{12} \approx -0.6559\} }\\
    p_{4,2}=\texttt{SDivide}(f_1,(x_1+\alpha_1),-\alpha_1), \ \ \ \texttt{RR}(p_{4,2})=\{\} \\
    p_{4,3}=\texttt{SDivide}(f_1,(x_1+\alpha_1),-\alpha_1), \ \ \ \texttt{RR}(p_{4,3})=\textcolor{blue}{\{\pm \beta_{13} \approx -0.6559 \} }\\
    \end{array}
\end{equation}
Note that $\beta_{12}$ is not equal to $\beta_{13}$, they differ after 10 significant figures. 
%
%
%
The full list of cells decomposing the second dimension is now:
%
%
\begin{equation}
  \begin{array}{l}
    Inequalities(\textbf{F}_3)=\\
    \{ \textcolor{blue}{b_{1,1}=\{x_2<-\beta_8\},b_{1,2}=\{x_2=-\beta_8\}}, b_{1,3}=\{-\beta_8<x_2<\beta_8\},\\ \textcolor{blue}{b_{1,4}=\{x_2=\beta_8\},
        ,b_{1,5}=\{\beta_8<x_2 \}}, \\ 
        \textcolor{blue}{b_{2,1}=\{x_2<-1\},b_{2,2}=\{x_2=-1\},b_{2,3}=\{-1<x_2<0\}},\\
        \textcolor{blue}{b_{2,4}=\{x_2=-0\},b_{2,5}=\{0<x_2<1 \},b_{2,6}=\{x_2=1 \},b_{2,7}=\{1<x_2\}},\\
         \textcolor{blue}{b_{3,1}=\{x_2<-\beta_{11}\},b_{3,2}=\{x_2=-\beta_{11}\},b_{3,3}=\{-\beta_{11}<x_2<-\beta_{10}\}},\\
         \textcolor{blue}{b_{3,4}=\{x_2=-\beta_{10}\},b_{3,5}=\{-\beta_{10}<x_2<\beta_{10} \},b_{3,6}=\{x_2=\beta_{10} \}},\\
         \textcolor{blue}{b_{3,7}=\{\beta_{10}<x_2<\beta_{11}\},b_{3,8}=\{x_2=\beta_{11} \},b_{3,9}=\{\beta_{11}<x_2\}},\\
         \textcolor{blue}{b_{4,1}=\{x_2<-\beta_{13}\},b_{4,2}=\{x_2=-\beta_{13}\},b_{4,3}=\{-\beta_{13}<x_2<-\beta_{12}\}},\\
         \textcolor{blue}{b_{4,4}=\{x_2=-\beta_{12}\},b_{4,5}=\{-\beta_{12}<x_2<\beta_{12} \},b_{4,6}=\{x_2=\beta_{12} \}},\\
         \textcolor{blue}{b_{4,7}=\{\beta_{12}<x_2<\beta_{13} \},b_{4,8}=\{x_2=\beta_{13} \},b_{4,9}=\{\beta_{13}<x_2\}},\\
         \textcolor{blue}{b_{5,1}=\{x_2<-\beta_0\},b_{5,2}=\{x_2=-\beta_0\},b_{5,3}=\{-\beta_0<x_2<-\beta_9\}},\\
        \textcolor{blue}{b_{5,4}=\{x_2=-\beta_9\},b_{5,5}=\{-\beta_9<x_2<\beta_9 \},b_{5,6}=\{x_2=\beta_9 \}},\\
        \textcolor{blue}{b_{5,7}=\{\beta_9<x_2<\beta_0\},b_{5,8}=\{x_2=\beta_0 \},b_{5,9}=\{\beta_0<x_2\}},\\
        b_{6,1}=\{x_2<-1\},b_{6,2}=\{x_2=-1\},\\
        b_{6,3}=\{-1<x_2<1\},b_{6,4}=\{x_2=1\},b_{6,5}=\{1<x_2\},\\
         b_{7,1}=\{x_2<-\beta_1\},b_{7,2}=\{x_2=-\beta_1\},b_{7,3}=\{-\beta_1<x_2<-\beta_2\},\\
         b_{7,4}=\{x_2=-\beta_2\},b_{7,5}=\{-\beta_2<x_2<\beta_2 \},b_{7,6}=\{x_2=\beta_2 \},\\
         b_{7,7}=\{\beta_2<x_2<\beta_1 \},b_{7,8}=\{x_2=\beta_1 \},b_{7,9}=\{\beta_1<x_2\},\\
         b_{8,1}=\{x_2<-\beta_3\},b_{8,2}=\{x_2=-\beta_3\},b_{8,3}=\{-\beta_3<x_2<-\beta_4\},\\
         b_{8,4}=\{x_2=-\beta_4\},b_{8,5}=\{-\beta_4<x_2<\beta_4 \},b_{8,6}=\{x_2=\beta_4 \},\\
         b_{8,7}=\{\beta_4<x_2<\beta_3 \},b_{8,8}=\{x_2=\beta_3 \},b_{8,9}=\{\beta_3<x_2\},\\
        b_{9,1}=\{x_2<-\beta_6\},b_{9,2}=\{x_2=-\beta_6\},b_{9,3}=\{-\beta_6<x_2<-\beta_5\},\\
        b_{9,4}=\{x_2=-\beta_5\},b_{9,5}=\{-\beta_5<x_2<\beta_5 \},b_{9,6}=\{x_2=\beta_5 \},\\
        b_{9,7}=\{\beta_5<x_2<\beta_6\},b_{9,8}=\{x_2=\beta_6 \},b_{9,9}=\{\beta_6<x_2\},\\
        b_{10,1}=\{x_2<-1\},b_{10,2}=\{x_2=-1\},b_{10,3}=\{-1<x_2<0\},\\
        b_{10,4}=\{x_2=-0\},b_{10,5}=\{0<x_2<1 \},b_{10,6}=\{x_2=1 \},b_{10,7}=\{1<x_2\},\\
        b_{11,1}=\{x_2<-\beta_7\},b_{11,2}=\{x_2=-\beta_7\},b_{11,3}=\{-\beta_7<x_2<\beta_7\},\\
        b_{11,4}=\{x_2=\beta_7\},b_{11,5}=\{\beta_7<x_2 \}
    \}
    \end{array}
\end{equation}

Figures 10-12 show the new CAD tree structure and its split into new and unchanged cells.
As you can see above, much of the CAD structure was able to be stored and reused drastically saving the solver from re-computation costs. 

\begin{figure}[p]
  \centering
  \begin{minipage}[b]{0.9\textwidth}
    \includegraphics[width=\textwidth]{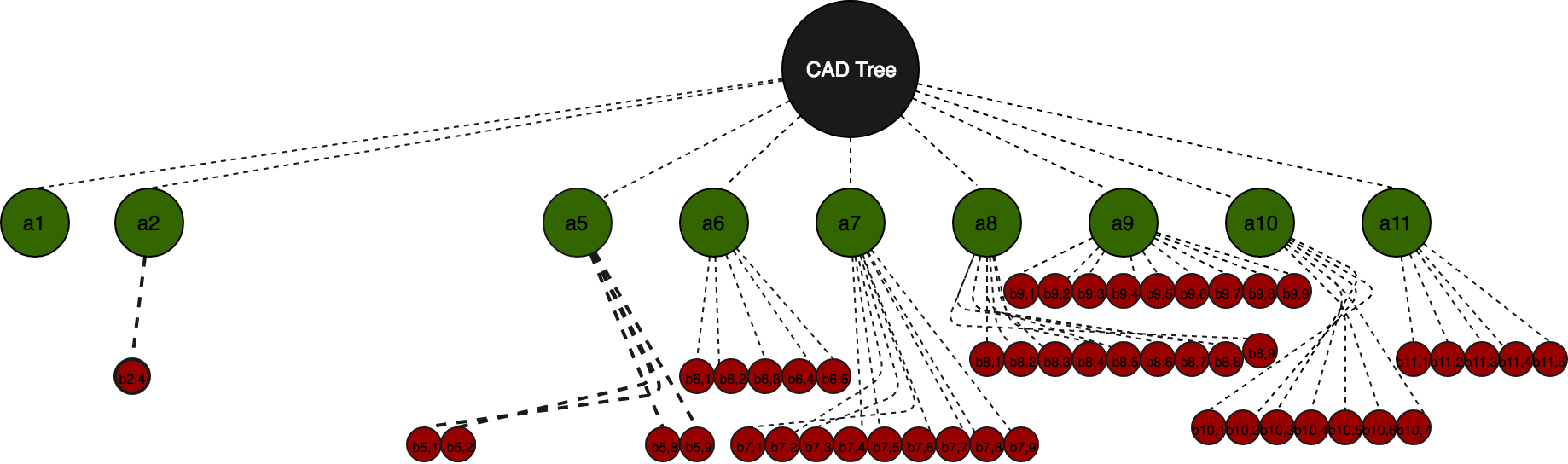}
    \caption{CAD tree of \textbf{unchanged cells} from $F_1$ incremented by $f_4$, forming the CAD tree to $F_3$, where x1=$x_1$ and x2=$x_2$. The green nodes represent cells in the first dimension and red nodes represent cells in the second dimension.}
  \end{minipage} 
\end{figure}

\begin{figure}[p]
  \centering
  \begin{minipage}[b]{0.9\textwidth}
    \includegraphics[width=\textwidth]{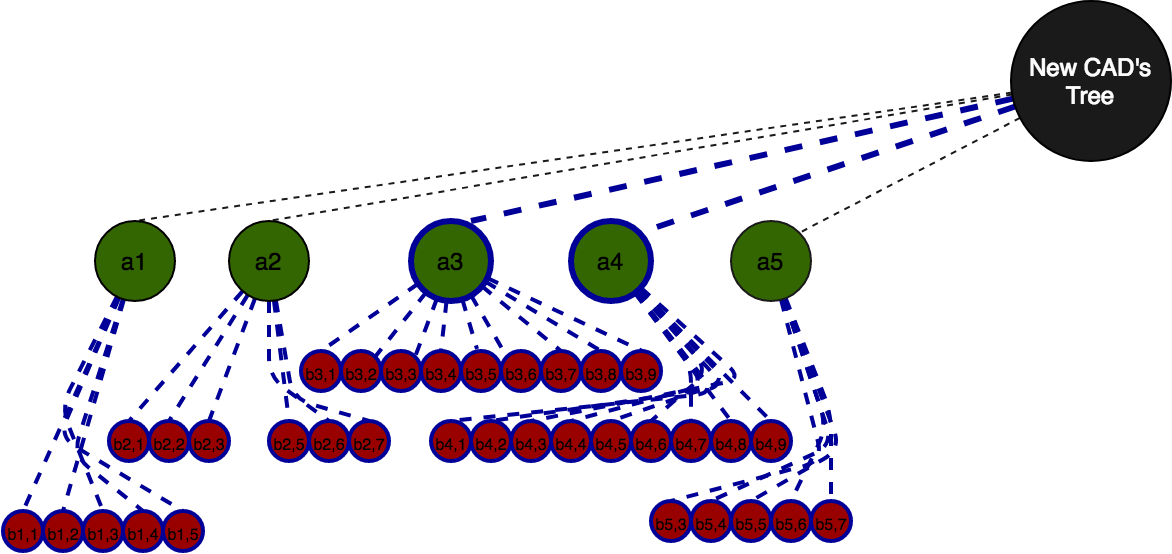}
    \caption{CAD tree \textbf{new cells} from $F_1$ incremented by $f_4$, forming the CAD tree to $F_3$, where x1=$x_1$ and x2=$x_2$. The green nodes represent cells in the first dimension and red nodes represent cells in the second dimension. Blue outlines around lines/ nodes represent new connections/ cells.}
  \end{minipage} 
\end{figure}

\begin{figure}[p]
  \centering
  \begin{minipage}[b]{0.9\textwidth}
    \includegraphics[width=\textwidth]{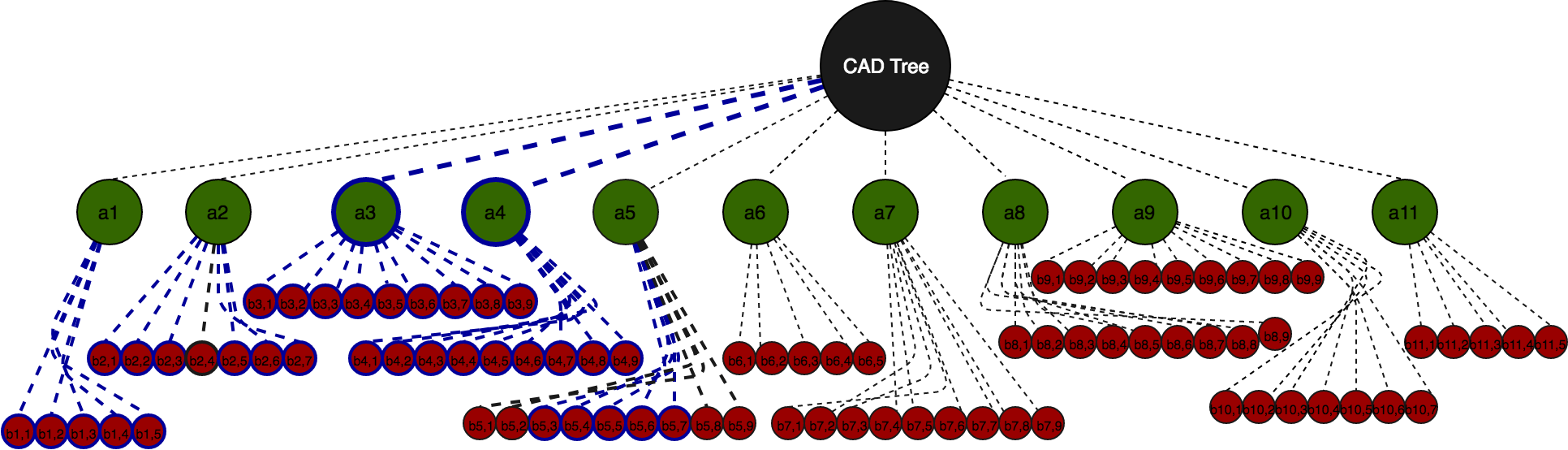}
    \caption{CAD tree of $F_1$ incremented by $f_4$, forming the CAD tree to $F_3$, where x1=$x_1$ and x2=$x_2$. The green nodes represent cells in the first dimension and red nodes represent cells in the second dimension. Blue outlines around lines/ nodes represent new connections/ cells.}
  \end{minipage} 
\end{figure}

\subsection{Algorithms}

When incrementing the lift stage, we can think of it as starting at the root node of the old CAD tree and working our way down it, one depth level at a time, until we reach the leaves. In the process of working down the old tree, we will be creating subtrees, which will later be reconnected to the unchanged tree, to form the new incremented CAD tree. 
\par 

One tree (UnchangedCads) is a strict subset of the old trees nodes and connections, discovered through following the old structure down and Lazard evaluating each cell not marked as "new", at each depth $p$'s with new projection polynomials in $R^{p+1}$. Then, if there are new roots, we prune inspected cells children, and the cell is sent to the NewCads set for full re-computation. In the NewCads set, we will then use this cell to form a subtree, to later be reconnected with the UnchangedCads tree. Each cell in the NewCads list is a subtree, to later all be reconnected via source indices. 
\par 

When going through the old CAD tree structure, we have only two cases: 
\par 

$\textbf{CASE 1:}$ When a node has new roots, and thus new children: 
\par 

Each new root has been acquired from one of the new projection polynomials. We start by pruning all of the child branches in the old tree structure, by labelling them as "new", then performing a full lift onto the set of all projection polynomials from $\mathbb{R}^{k}, \ldots, \mathbb{R}^{n}$, where $k$ is the depth the new root was discovered. We separate these cells into a list called newcad. When we label a cell as new, effectively that halts tree growth in the UnchangedCads structure, so that later on we can just attach the branch extensions, gained from the incremental lift.
\par 

When moving a cell into newcad, we make sure the cell has an updated source index, as its source cell would now be saved in a new tree structure with a new index. 
\par 

$\textbf{CASE 2:}$ When a node has no new roots,
\par 

While going through the old tree, we check the source cell of each cell not labelled "new", to see if its parent was tagged as new, if so we mark it as "new". Otherwise, we label it "old", thus pass it into UnchangedCads. UnchangedCads is where we save the cost from unnecessary calculations, as cells in this tree are only Lazard evaluated at each new projection polynomial rather than all projection polynomials. If they have a fresh root, we move it over to the NewCads structure. Otherwise, we continue to attach it's child cells and their branches (source cell index). 
\par 

When moving cells into UnchangedCads, we make sure that the indexing of each cell does not clash with that of the indexing in NewCads at each depth level in the tree. We then merge the NewCads and UnchangedCads trees, forming the full incremented CAD tree. At each stage of the lift we merge-sort the list of cells, and because of the way be indexed them the source indices don't change. 
\par 

The main difficulties with implementing this method were: 
\par 

\begin{enumerate} 
\item A cell creation memory system to trace through which cell in, say $\mathbb{R}^{3}$, were created by a cell in $\mathbb{R}^{2}$. Solved through a combination of the structure outlined below.
\item A function which could split cells in the first variable based on new roots appearing. We created the \texttt{LiftSplit} function for this, 
\item We would need to store the real roots in the first variable of a CAD to search through the cells efficiently. Created a list passed through the lift functions which stores a sorted and unique list of roots in the first variable. 
\item A CAD $\mathbb{R}^{1}$ and  $\mathbb{R}^{1}$ sort and relabelling function to ensure the roots of the first variable are sorted in ascending order for optimised multiple applications of incremental lift. We created the function \texttt{Combine} to do this at the very end of the lifting process.
\item We would need a way of determining when a CAD cell at a higher stage has not been split at a lower stage, whether its source cell has been saved in the new tree structure and if so what the new source index is.
\item Thinking of cases of empty projection cells, how to use them to our advantage. We had to create many logical gates to work with these cases as well as creating an optimisation function \texttt{LiftAddOptimised}. This optimisation function activates if the first number of new projection polynomials sets were empty, it would copy the old tree structure down to a depth until it encounters a new root obtained from the new projection polynomials. We then begin the incremental algorithm from this boosted depth. 
\end{enumerate}

We outlined the main issues above, and then we describe the steps we took to overcome them. We will introduce you to the new structure of a cad cell before moving through the details. A cell can be described now by 

\begin{equation}
\textnormal{[[index],[constraints],[sample points],[source cell index],[split branch flag]]}
\end{equation} 
\begin{equation}
\textnormal{e.g.} [[3],[-1<x<1,y<1],[0,0],[2],["new"]],
\end{equation}
For example, during the incremental CAD example we saw the cell (\ref{cell1}) below. This structure informs the system that cell [2] was split by a new root in $x$, or lifted onto the old and new roots in $y$. 
\begin{equation}
\label{cell1}
[[2],[-1<x<1],[0],[1],["new"]].
\end{equation}  

The next example cell (\ref{cell2}) tells us the new index of cell 3 in the second depth of the old cad tree, has now changed to four. So when we later move on to cell 3's children, we can adjust there source indices based on the new tree structure, to ensure the subtrees are connected correctly. 
\begin{equation}
\label{cell2}
[[3],[-1<x<1,y<1],[0,0],[2],[4]].
\end{equation}  

\begin{algorithm}
\caption{LiftSetup}
\label{alg:6}
{\fontsize{10}{15}\selectfont
\begin{algorithmic}[1]
\STATE \textbf{Input:} A table of sets of projection polynomials $pset= \{p_1,\ldots,p_m\} \in \mathbb{R}[x_1,\ldots,x_n]$, $vars=[x_{1},\ldots,x_n]$ 
\STATE \textbf{Output:} A CAD $ cad_1 = \ [{cell1_1,\ldots,cell1_q}] \in \mathbb{R}[x_1] $ in the last variable $x_n$, with information ($LiftInf$) to construct the cells in the second last variable ($x_2$). 
\STATE \textbf{Procedure} \texttt{Lazard LiftSetup}
\STATE $cad \leftarrow table():$
\STATE $roots \leftarrow []:$
\STATE $LiftInf_2 \leftarrow []:$ \COMMENT{where $LiftInf_i$ in the information required to lift CAD in $\mathbb{R}^{i-1}$ to a CAD in $\mathbb{R}^{i}$}
\FOR{\texttt{$i$ from $1$ to $n\_elements(pset[1])$}}
\STATE $roots \leftarrow$ Concatenate $roots$ with the real roots from $pset[1][i] \in \mathbb{R}[x_1]$
\ENDFOR 
\STATE $roots \leftarrow$ Sort and remove duplicates from $roots$ 
\STATE $cad_1 \leftarrow$  Generate the CAD cells in the first variable $x_1$ from $roots$
\FOR{\texttt{$i$ from $1$ to $n\_elements(cad_1)$}}
\FOR{\texttt{$j$ from $1$ to $n\_elements(pset[2])$}}
\STATE $roots \leftarrow$ \texttt{Lazard\_evaluation}$(cad_1[i],pset[2][j])$ 
\STATE Note: $pset[2][j] \in \mathbb{R}[x_n,x_2]$
\IF{\texttt{roots<>[]}}
\STATE $LiftInf_2 \leftarrow [LiftInf_2,[[i],[roots]]]$
\STATE Where i is the cell index (note this is not the CAD index, this could be easily implemented in place, however for ease of iteration we chose to use a scalar index) and roots is a unique list of roots in $x_2$ for the cell $c_i$.
\ENDIF
\ENDFOR 
\ENDFOR \COMMENT{We now have all the required information for lifting to $x_2$}
\STATE \textbf{return} \(cad_1,LiftInf_2\)
\end{algorithmic}
}
\end{algorithm}

\begin{algorithm}
\caption{Lift}
\label{alg:8}
{\fontsize{10}{15}\selectfont
\begin{algorithmic}[1]
\STATE \textbf{Input:} A table of sets of projection polynomials $pset= \{p_1,\ldots,p_m\} \in \mathbb{R}[x_1,\ldots,x_n]$, $vars=[x_{1},\ldots,x_n]$ 
\STATE \textbf{Output:} A CAD \( cad_n = \{celln_1,\ldots,celln_q\} \in \mathbb{R}^n \)
\STATE \textbf{Procedure} \texttt{Lazard Lift}
\STATE $dim \leftarrow n\_elements(vars)$
\STATE $cad_1,LiftInf_2 \leftarrow \texttt{LiftSetup}(pset,vars)$
\FOR{\texttt{$k$ from $2$ to $dim$}} 
\STATE $cad_k \leftarrow []:$
\STATE $LiftInf_{k+1} \leftarrow []:$
\STATE $cad_k \leftarrow LIFT(cad_{k-1},LiftInf_k$ ) \STATE \COMMENT{Note: This lift is just the classical collins lift operation, however we are using the Lazard operators minimal polynomial set}
\IF{\texttt{k<>dim}}
\FOR{\texttt{$i$ from $1$ to $n\_elements(cad_k)$}}
\FOR{\texttt{$j$ from $1$ to $n\_elements(pset[k+1])$}}
\STATE $roots \leftarrow \texttt{Lazard\_evaluation}(cad_k[i],pset[K+1][j])$ 
\STATE \COMMENT{$pset[2][j] \in \mathbb{R}[x_n,x_2]$}
\IF{\texttt{roots<>[]}}
\STATE $LiftInf_{k+1} \leftarrow [LiftInf_{k+1},[[i],[roots]]]$
\STATE \COMMENT{Where i is the cell index and roots is a unique list of roots in $x_2$ for the cell $c_i$.}
\ENDIF  
\ENDFOR
\ENDFOR 
\ENDIF  
\ENDFOR
\STATE \textbf{return} \(cad_{dim}\)
\end{algorithmic}
}
\end{algorithm}

\begin{algorithm}
\caption{LiftSetupAdd}
\label{alg:10}
{\fontsize{10}{15}\selectfont
\begin{algorithmic}[1]
\STATE \textbf{Input:} A sets of new projection polynomials $psetnew$, a table of sets of all  projection polynomials $psetfull$, $vars$ and previous cad cells.
\STATE \textbf{Output:} A list [NewCads, OldCad, OldRoots, UnchangedCads].  NewCads in the last variable $x_n$, 
OldRoots of all real roots in last variable,OldCad which contains the previous structure, UnchangedCads which is the copied and unchanged old structure.
\STATE \textbf{Procedure} \texttt{Lazard LiftSetup}
\STATE $dim \leftarrow$ Number of elements in $vars$
\STATE $cad \leftarrow table()$
\STATE $NewRoots \leftarrow []$
\STATE $NewCads \leftarrow table()$
\STATE $UnchangedCads \leftarrow table()$
\STATE $LiftInf_2 \leftarrow []$ 
\FOR{\texttt{$i$ from $1$ to $n\_elements(psetnew[1])$}}
\STATE $NewRoots \leftarrow [NewRoots,RealRoots(psetnew[1][i])]$  
\STATE $NewRoots \leftarrow$ Sort in ascending order and remove duplicates of $NewRoots$ 
\ENDFOR \textcolor{blue}{
\STATE $NewCads[1],UnchangedCads[1]=\newline
\texttt{SplitCells}(OldRoots,NewRoots,OldCad)$
\COMMENT{\textbf{Adjustment 1}}
\FOR{\texttt{$i$ from $1$ to $n\_elements(oldcad[1])$}}
\FOR{\texttt{$j$ from $1$ to $n\_elements(psetnew[2])$}}
\IF{$0<>$\texttt{Lazard\_evaluation}$(oldcad[1][i],psetnew[2][j])$} 
\STATE $NewCads \leftarrow $ Concatenate $oldcad[1][i]$  to $NewCads$
\ENDIF
\ENDFOR 
\ENDFOR 
\FOR{\texttt{$i$ from $1$ to $n\_elements(NewCads[1])$}}
\FOR{\texttt{$j$ from $1$ to $n\_elements(psetfull[2])$}}
\STATE $roots$=\texttt{Lazard\_evaluation}$(oldcad[1][i],psetnew[2][j]$
\STATE $LiftInf_2 \leftarrow [LiftInf_2,[[i],[roots]]]$
\ENDFOR 
\ENDFOR 
\FOR{\texttt{$i$ from $1$ to $n\_elements(oldcad[1])$}}
\STATE \textbf{If} cell $OldCad[1][i]$'s flag is not equal to "new", then concat cell to $UnchangedCads[1]$ and update index accordingly. 
\ENDFOR }
\STATE $OldRoots$= Merge $OldRoots$ and $NewRoots$
\COMMENT{\textbf{Adjustment 3}}
\STATE \textbf{return} list of outputs
\end{algorithmic}
}
\end{algorithm}

\begin{algorithm}
\caption{LiftAdd}
\label{alg:liftadd}
{\fontsize{10}{15}\selectfont
\begin{algorithmic}[1]
\STATE \textbf{Input:} A sets of new projection polynomials $psetnew$, a table of sets of all  projection polynomials $psetfull$, $vars$ and previous cad cells OldCad.
\STATE \textbf{Output:} Full projection set psetfull, InCAD
\STATE \textbf{Procedure} \texttt{Lazard LiftAdd}
\STATE $[NewCads, OldCad, OldRoots, UnchangedCads] \leftarrow \texttt{LiftSetupAdd}(pset,vars)$
\STATE $dim \leftarrow$ Number of elements in $vars$
\textcolor{blue}{
\FOR{\texttt{$d$ from $2$ to $dim-1$}} 
\STATE $LiftInf_{d+1}\leftarrow []$ 
\STATE $NewCads[d] \leftarrow LIFT(NewCads_{d-1},LiftInf_d$ ) 
\FOR{\texttt{$i$ from $1$ to $n\_elements(OldCad[d])$}}
\FOR{\texttt{$j$ from $1$ to $n\_elements(psetnew[d+1])$}}
\IF{$0<>$\texttt{Lazard\_evaluation}$(OldCad[d][i],psetnew[d+1][j])$} 
\STATE $NewCads \leftarrow $ Concatenate $oldcad[d][i]$  to $NewCads$
\ENDIF
\ENDFOR 
\ENDFOR 
\FOR{\texttt{$i$ from $1$ to $n\_elements(NewCads[d])$}}
\FOR{\texttt{$j$ from $1$ to $n\_elements(psetfull[d+1])$}}
\STATE $roots$=\texttt{Lazard\_evaluation}$(OldCad[d][i],psetnew[d+1][j]$
\STATE $LiftInf_{d+1} \leftarrow [LiftInf_{d+1},[[i],[roots]]]$
\ENDFOR 
\ENDFOR 
\FOR{\texttt{$i$ from $1$ to $n\_elements(OldCad[d])$}}
\STATE \textbf{If} cell $OldCad[d][i]$'s flag is not equal to "new", then concat cell to $UnchangedCads[d]$ and update index accordingly. 
\ENDFOR 
\ENDFOR 
\STATE $FinalUnchangedCADS \leftarrow []$
\STATE $FinalUnchangedCADS$ Union $UnchangedCads[d][f]$ for all cells $f$ with their corresponding flags not equal to "new". 
\STATE $InCAD \leftarrow FinalUnchangedCADS$ Union $NewCads[d]$ }
\STATE \textbf{return} psetfull, InCAD
\end{algorithmic}
}
\end{algorithm}

The algorithm for the Lazard Incremental Lift adjusted as follows:
\begin{enumerate}
\item First, it lifts the NewCads based on the previously obtained information. UnchangedCads
\item Then it checks to see which cells from old CAD to move into NewCads for a further lift or to copy structure over to UnchangedCads while preserving root order.
\item Gathers the necessary lift information on NewCads, then repeats.
\end{enumerate}

\subsection{Experiments: Incremental lifting}
\par 

We conducted testing for the incremental lift method on the same examples used to test the projection methods earlier.

\subsubsection{Experiment: Tri-variate}
The majority of cases were faster, on average faster by $17\%$, but there were examples up to $28\%$ slower in. The net speed difference was $7\%$ increase in performance with the incremental lift.
\par

\begin{center}
\begin{tabular}{@{}*4l@{}}
  \toprule[1.5pt]
  \multicolumn{1}{c}{\head{Lift}} &
    \multicolumn{2}{c}{\head{Results}} \\ 
  \head{} & \head{Classical} & \head{Incremental} & \head{}\\
  \cmidrule(l){1-1}\cmidrule(r){2-4}
  \verb|Variance| & \rmfamily 0.05541s & \rmfamily  0.06838s & \rmfamily  \textcolor{red}{\textbf{18.96\%}} Larger\\
  \verb|Mean| & \rmfamily 0.2880s & \rmfamily  0.2687s & \rmfamily  \textcolor{green}{\textbf{6.707\%}} Faster \\
  \verb|Lower Quartile| & \rmfamily 0.1275s & \rmfamily  0.0995s & \rmfamily  \textcolor{green}{\textbf{21.96\%}} Faster \\
  \verb|Median| & \rmfamily 0.207s & \rmfamily  0.164s & \rmfamily  \textcolor{green}{\textbf{20.77\%}} Faster\\
  \verb|Upper Quartile| & \rmfamily 0.3605s & \rmfamily 0.3523s & \rmfamily  \textcolor{green}{\textbf{2.29\%}} Faster \\
  \bottomrule[1.5pt]
\end{tabular}
\end{center}

\subsubsection{Experiment: Bi-variate}
When looking at the cases which were faster, they were on average faster by $31\%$, and $13\%$ slower in the other case. The net speed difference was a $30\%$ increase in performance with the incremental lift.
\par 

\begin{center}
\begin{tabular}{@{}*4l@{}}
  \toprule[1.5pt]
  \multicolumn{1}{c}{\head{Lift}} &
    \multicolumn{2}{c}{\head{Results}} \\ 
  \head{} & \head{Classical} & \head{Incremental} & \head{}\\
  \cmidrule(l){1-1}\cmidrule(r){2-4}
  \verb|Variance| & \rmfamily 0.003734s & \rmfamily  0.002903s & \rmfamily  \textcolor{green}{\textbf{28.63\%}} Smaller\\
  \verb|Mean| & \rmfamily 0.1778s & \rmfamily  0.1240s & \rmfamily  \textcolor{green}{\textbf{30.25\%}} Faster \\
  \verb|Lower Quartile| & \rmfamily 0.1328s & \rmfamily  0.089s & \rmfamily  \textcolor{green}{\textbf{32.96\%}} Faster \\
  \verb|Median| & \rmfamily 0.163s & \rmfamily  0.1255s & \rmfamily  \textcolor{green}{\textbf{23.01\%}} Faster\\
  \verb|Upper Quartile| & \rmfamily 0.226s & \rmfamily  0.163s & \rmfamily  \textcolor{green}{\textbf{27.87\%}} Faster \\
  \bottomrule[1.5pt]
\end{tabular}
\end{center}

\subsection{Experiments: Full incremental CAD}

The complete Incremental CAD system consists only of the incremental projection and incremental lift functions combined consecutively as one function.

\subsubsection{Experiment: Tri-variate}
When looking at the cases which were faster, they were on average faster by $19\%$, and $33\%$ slower in the other cases. The net speed difference was a $12\%$ increase in performance with the full incremental CAD across all cases.
\par

\begin{center}
\begin{tabular}{@{}*4l@{}}
  \toprule[1.5pt]
  \multicolumn{1}{c}{\head{CAD}} &
    \multicolumn{2}{c}{\head{Results}} \\ 
  \head{} & \head{Classical} & \head{Incremental} & \head{}\\
  \cmidrule(l){1-1}\cmidrule(r){2-4}
  \verb|Variance| & \rmfamily 0.05880s & \rmfamily  0.08517s & \rmfamily  \textcolor{red}{\textbf{30.96\%}} Larger\\
  \verb|Mean| & \rmfamily 0.3296s & \rmfamily  0.2908s & \rmfamily  \textcolor{green}{\textbf{11.79\%}} Faster \\
  \verb|Lower Quartile| & \rmfamily 0.174s & \rmfamily  0.1305s & \rmfamily  \textcolor{green}{\textbf{25.0\%}} Faster \\
  \verb|Median| & \rmfamily 0.24s & \rmfamily  0.182s & \rmfamily  \textcolor{green}{\textbf{24.17\%}} Faster\\
  \verb|Upper Quartile| & \rmfamily 0.411s & \rmfamily  0.3435s & \rmfamily  \textcolor{green}{\textbf{16.43\%}} Faster \\
  \bottomrule[1.5pt]
\end{tabular}
\end{center}

\subsubsection{Experiment: Bi-variate}
When looking at the cases which were faster, they were on average faster by $38\%$. The net speed difference was thus a $38\%$ increase in performance with the full incremental CAD across all cases.

\begin{center}
\begin{tabular}{@{}*4l@{}}
  \toprule[1.5pt]
  \multicolumn{1}{c}{\head{CAD}} &
    \multicolumn{2}{c}{\head{Results}} \\ 
  \head{} & \head{Classical} & \head{Incremental} & \head{}\\
  \cmidrule(l){1-1}\cmidrule(r){2-4}
  \verb|Variance| & \rmfamily 0.003896s & \rmfamily  0.002710s & \rmfamily  \textcolor{green}{\textbf{43.74\%}} Smaller\\
  \verb|Mean| & \rmfamily 0.2089s & \rmfamily  0.1314s & \rmfamily \textcolor{green}{ \textbf{37.12\%}} Faster \\
  \verb|Lower Quartile| & \rmfamily 0.153s & \rmfamily  0.101s & \rmfamily  \textcolor{green}{\textbf{33.99\%}} Faster \\
  \verb|Median| & \rmfamily 0.22s & \rmfamily  0.1305s & \rmfamily  \textcolor{green}{\textbf{40.68\%} }Faster\\
  \verb|Upper Quartile| & \rmfamily 0.2503s & \rmfamily  0.1633s & \rmfamily  \textcolor{green}{\textbf{34.77\%}} Faster \\
  \bottomrule[1.5pt]
\end{tabular}
\end{center}

We think the reason for such dramatic drops in performance in more complicated cases for the lifting stage, was due to poor choices of Maple's data-structure: in particular Maple lists which are implemented as immutable types meaning our edits of them caused separate lists to be created each time.   

\section{Further work}

Our implementation needs to be refactored into more appropriate data structures as discussed above.  
The next step after that would be to implement a reduction CAD system (where instead of incrementally adding polynomials we can remove them). To remove a polynomial from the CAD results in finding all those projection polynomials created from purely that source polynomial as well as resultants with only the source polynomial and another polynomial and removing them and their cad cells from the system. You would then need to merge the cells which had neighbouring cells removed, possibly having to perform a full CAD on the source cells affected. The difficulty is keeping track of the fact that single projection polynomials can be computed in different ways and so a full trace would need to be maintained.

We only explored systems in which we assumed every variable in the incremental polynomial, was already represented in the system. However, situations could arise when you need another dynamic level to the system, one in which the incrementing polynomial contains at least one unseen variable $x_{n+1}$, this would require a lot of recalculations, however, would be able to benefit from storing previous calculations, as seen with this proof of concept.



\section*{Acknowledgements}

This work described was funded by the European Union's Horizon 2020 research and innovation programme under grant agreement No H2020-FETOPEN-2015-CSA 712689 (\scsc).
We thank Chris Brown for an informal tutorial on the Lazard evaluation and the organisers of the \scsc 2017 Summer School at the Max Planck Institute for Informatics in Saarbr\"{u}cken, where this took place. We also thank Sam Timms, James Davenport and Stephen Forrest for useful discussions at the Summer School.

\bibliographystyle{splncs03}
\bibliography{references}

\section{Appendix}

\section*{Lazard Valuation \& Lazard Evaluation}

We will guide the reader through various Lazard lift examples, similar to those mentioned during Chris Brown's informal tutorial at the \scsc 2017 Summer School. We find these worked examples aid understanding the Lazard Lift process, crucial to implementing this article's contribution. 

\textbf{Background}

Brown's projection operator states that all you need is resultants, discriminants, and leading coefficients. However, it knows it may go wrong if a polynomial is nullified. This means that substituting for some but not all of the variables makes the polynomial vanish. E.g.
\begin{align}
xz^2+zy+y
\end{align}
is nullified at (x, y) = (0, 0). The problem is that there may be important information needed for delineability buried behind the thing that vanished. If you lose that information you miss real roots that should have been isolated. Actually, the example above is not too bad as the nullification happens only at one point. But if it happens over a region with dimension greater than zero it really does matter.

As previously discussed, Lazard's operator has resultants, discriminants, leading coefficients and trailing coefficients. So slightly bigger than Brown but no nullification problems.

Lazard's projection depends neither on sign-invariance or order-invariance. The proof is completely different as it relies on Lazard valuation invariance.

Think about evaluating a polynomial at a point. You can evaluate variables in any order you want. Lazard's valuation is more structured and you do not get a number out. 

Given a polynomial $f$ at $(x_1,x_2,\cdots,x_k)=(\alpha_1,\alpha_2,\cdots,\alpha_k)$. We will monitor two items:
\begin{enumerate}
\item A tuple of integers. Starting with the empty tuple (). 
\item A polynomial, starting with $f$. 
\end{enumerate}
Now that we have ((),f), we may begin. We will gradually add integers into this tuple and change the polynomial stored over time. 

\textbf{Basic Example}

$f=x+2y$ with $\alpha=(3,2)$. Start off by substituting in $x=3$:  Does the polynomial disappear? The answer is no, so we compute.
\begin{align}
((0),3+2y)
\end{align}
Does plugging in the $y = 2$ cause a zero? No, so we now have.
\begin{align}
((0, 0), 7)
\end{align}
Adding the zero to the tuple means not nullification.

\textbf{Intermediate Example}

$f=(x-2)y+(x-2)$ with $\alpha=(2,1)$. Does substituting $x=2$ nullify the polynomial? Yes! So now we need to keep dividing out by $x - 2$ until it does not vanish. Keep track of how many times you divided in the tuple. Only then do you continue to substitute values (if there is any left).  In this case we track:
\begin{align}
&((), (x-2)y+(x-2)) \nonumber \\
&((1), y+1)
\end{align}

\textbf{Second Intermediate Example}

$f=(x-2)y+x^2-2x$ with $\alpha=(2,1)$.  So this time:
\begin{align}
&((), (x-2)y+x^2-2x)) \nonumber \\
&((1), y+x)
\end{align}
This time there is an $x$ left. So now we can substitute:
\begin{align}
((1), y+2)
\end{align}
Now does it vanish at $y = 1$? Answer is no, so just substitute it in:
\begin{align}
((1,0), 3)
\end{align}
The item left over in the first tuple is what is referred to as Lazard's valuation.

\textbf{Non Trivial Example}

$f=(y^2-2y+x)(z-2)w^2-z(z+x-y-2)(xy-1)^2$. Variable order of $x,y,z,w$ with $\alpha=(1,1,2,0)$.  So we start with:
\begin{align}
((), (y^2-2y+x)(z-2)w^2-z(z+x-y-2)(xy-1)^2)
\end{align}
and substitute for $x=1$@
\begin{align}
((0), (y^2-2y+1)(z-2)w^2-z(z-y-1)(y-1)^2)
\end{align}
If we evaluate at $y=1$ the polynomial would vanish! Because first factor is $(y-1)^2$. We factor $y-1$ out twice. 
\begin{align}
((0,2), (z-2)w^2-z(z-y-1)) \nonumber \\
((0,2), (z-2)w^2-z(z-2))
\end{align}
If we evaluate at $z=2$ the polynomial is nullified.  So again divide:
\begin{align}
((0,2,1), w^2-z) \nonumber \\
((0,2,1), w^2-2)
\end{align}
Finally we substitute $w=0$ and end up with:
\begin{align}
((0,2,1,0), -2)
\end{align}

\textbf{Important Example}

Suppose we have a projection factor set:
${(y^2-2)w+z(y-x^2+x+2)}$  
and our sample points are $\alpha=(\sqrt{2},-\sqrt{2},1)$. Let us do Lazard valuation with respect to variable ordering $x,y,z,w$
\begin{align}
((), (y^2-2)w+z(y-x^2+x+2))
\end{align}
\begin{align}
((0), (y^2-2)w+z(y+\sqrt{2}))
\end{align}
\begin{align}
((0,1), (y-\sqrt{2})w+z) \nonumber \\
((0,1), -2 \sqrt{2}w+1)
\end{align}
\begin{align}
((0,1,1), -2 \sqrt{2}w+1)
\end{align}
This univariate polynomial needs to have its real root isolated. The real root isolation would preserve the invariance of Lazard valuation. 

\end{document}